\documentclass[
 reprint,
 amsmath,amssymb,
 aps,
 pra,
 superscriptaddress
]{revtex4-2}

\usepackage{graphicx}% Include figure files
\usepackage{dcolumn}% Align table columns on decimal point
\usepackage{bm}% bold math
\usepackage{braket}
\usepackage{CJK}
\usepackage{color}
\usepackage[
    hyperindex,
    breaklinks,
    colorlinks=true,
    linkcolor=blue,
    anchorcolor=red,
    citecolor=blue,
    urlcolor=blue
]{hyperref}
\newcommand{\poincare}[0]{Poincar\'e~}

\begin{document}

\preprint{APS/123-QED}

\title{Self-Similarity Among Energy Eigenstates}

\begin{CJK}{UTF8}{gbsn}
\author{Zhelun Zhang(张哲伦)}
\altaffiliation{These two authors contributed equally to this work.}
\affiliation{International Center for Quantum Materials, School of Physics, Peking University, 100871, Beijing, China}

\author{Zhenduo Wang(王朕铎)}
\altaffiliation{These two authors contributed equally to this work.}
\affiliation{International Center for Quantum Materials, School of Physics, Peking University, 100871, Beijing, China}

\author{Biao Wu(吴飙)}
\affiliation{International Center for Quantum Materials, School of Physics, Peking University, 100871, Beijing, China}
\affiliation{Wilczek Quantum Center, School of Physics and Astronomy, Shanghai Jiao Tong University, Shanghai 200240, China}
\affiliation{Collaborative Innovation Center of Quantum Matter, Beijing 100871,  China}

\begin{abstract}
In a quantum system, different energy eigenstates have different properties or features, allowing us
define a classifier to divide them into different groups. We find that the ratio of each type of energy eigenstates 
in an energy shell $[E_{c}-\Delta E/2,E_{c}+\Delta E/2]$ is invariant with changing width $\Delta E$ or 
 Planck constant $\hbar$ as long as the number of eigenstates in the shell is statistically large enough. 
 We give an argument that such self-similarity in energy eigenstates  is a general feature for all quantum systems, 
 which is further illustrated numerically with various quantum  systems, including circular billiard, 
 double top model, kicked rotor, and Heisenberg XXZ model. 
\end{abstract}

\maketitle
\end{CJK}

\section{Introduction} 
Energy eigenvalues and eigenstates are constitutive to the properties of quantum systems. 
They have already been thoroughly studied in various aspects. For eigenvalues, the well-known results include 
the Weyl law\cite{original_Weyl} and its generalization\cite{fractal_Weyl,Shepelyansky,Ramilowski,Pedrosa,multifractal_measures,Wiersig,Kopp,Korber,Eberspacher}, 
which describes the asymptotic behavior of the number of energy eigenvalues below an increasing energy. 
The distribution of nearest energy level spacings\cite{RevModPhys,Montambaux,signatures_chaos,Schagrin,spacing_ratio,Wang,Rao} 
is now widely used to characterize quantum systems: Wigner-Dyson distribution for chaotic systems and Poisson distribution for integrable systems. The degeneracy in eigen-energies and  their differences has been shown to be related to ergodicity and mixing 
in quantum systems\cite{Neumann1929,vonNeumann2010,qmixing}.

For energy eigenstates, there are also  many interesting  results.
These earliest studies have focused on the correlation and amplitude distribution of a single energy eigenstate\cite{RevModPhys,AURICH1993185,Backer,Xiong,Beugeling,Clauss,Keating,McDonald}. 
This line of studies ultimately leads to a well known hypothesis by Berry: each energy eigenstate has a Wigner function concentrated on the region explored by a typical orbit over infinite times in the semiclassical limit; or, equivalently, each energy eigenstate becomes a minimal invariant ensemble distribution in classical phase space in the semiclassical limit\cite{Berry_1977}. 
%This hypothesis is the basis of the well known eigenstate thermalization hypothesis (ETH).  
Recently, there have been studies on the single energy eigenstate in spin systems, which have no well defined semiclassical 
limit\cite{tight-banding,AAmodel,MBL_mutifractal,Tikhonov,Santos,qft,RM_fractal}.

In this work we focus on a sequence of energy eigenstates in an energy shell 
${\mathcal S}(E_c, \Delta E)=[E_{c}-\Delta E/2,E_{c}+\Delta E/2]$.  As these eigenstates have different physical 
properties or features, we define a classifier for a given physical property or feature and divide these eigenstates into different groups. 
We find  self-similarity among energy eigenstates for all quantum systems in the following sense:
if the ratio of the energy eigenstates having property $A$ is $f$ in the energy shell ${\mathcal S}(E_c, \Delta E)$,  
the ratio is still $f$ in the sub-shell ${\mathcal S}(E_c', \Delta E')\subset {\mathcal S}(E_c, \Delta E)$ as long as
the number of eigenstates in the sub-shell is statistically large enough. The self-similarity is particularly pronounced 
in the semiclassical limit $\hbar\rightarrow 0$, where the number of eigenstates in a very narrow energy shell is very large. 

We first illustrate such  self-similarity with a simple model, circular billiard with analytical results and extensive 
numerical computation.  We then give an analysis, arguing that such  self-similarity is generic feature for 
any quantum system that has a well-defined semiclassical limit. We finally illustrate the self-similarity with more 
examples, which include coupled tops, kicked rotor, and Heisenberg XXZ model. The result for the XXZ model 
is of particular interest as it shows that the self-similarity exists even in quantum systems that have no
well-defined semiclassical limits. In the end, we argue that such a self-similarity offers a good explanation 
why the microcanonical ensemble in quantum statistical mechanics, which is established on the equal probability hypothesis,  
works for all quantum systems regardless of their integrability.
%As a result, ETH is just a special case of the microcanonical ensemble.

\section{Self-similarity in Energy Eigenstates}\label{sec:theory}
Before general discussion, we study a simple but illustrative example, a quantum circular billiard\cite{circular}, where the self-similarity in its energy eigenstates can be demonstrated convincingly through analysis and extensive numerical calculation. 

\begin{figure}[h]
    \centering
    \includegraphics[width=0.85\linewidth]{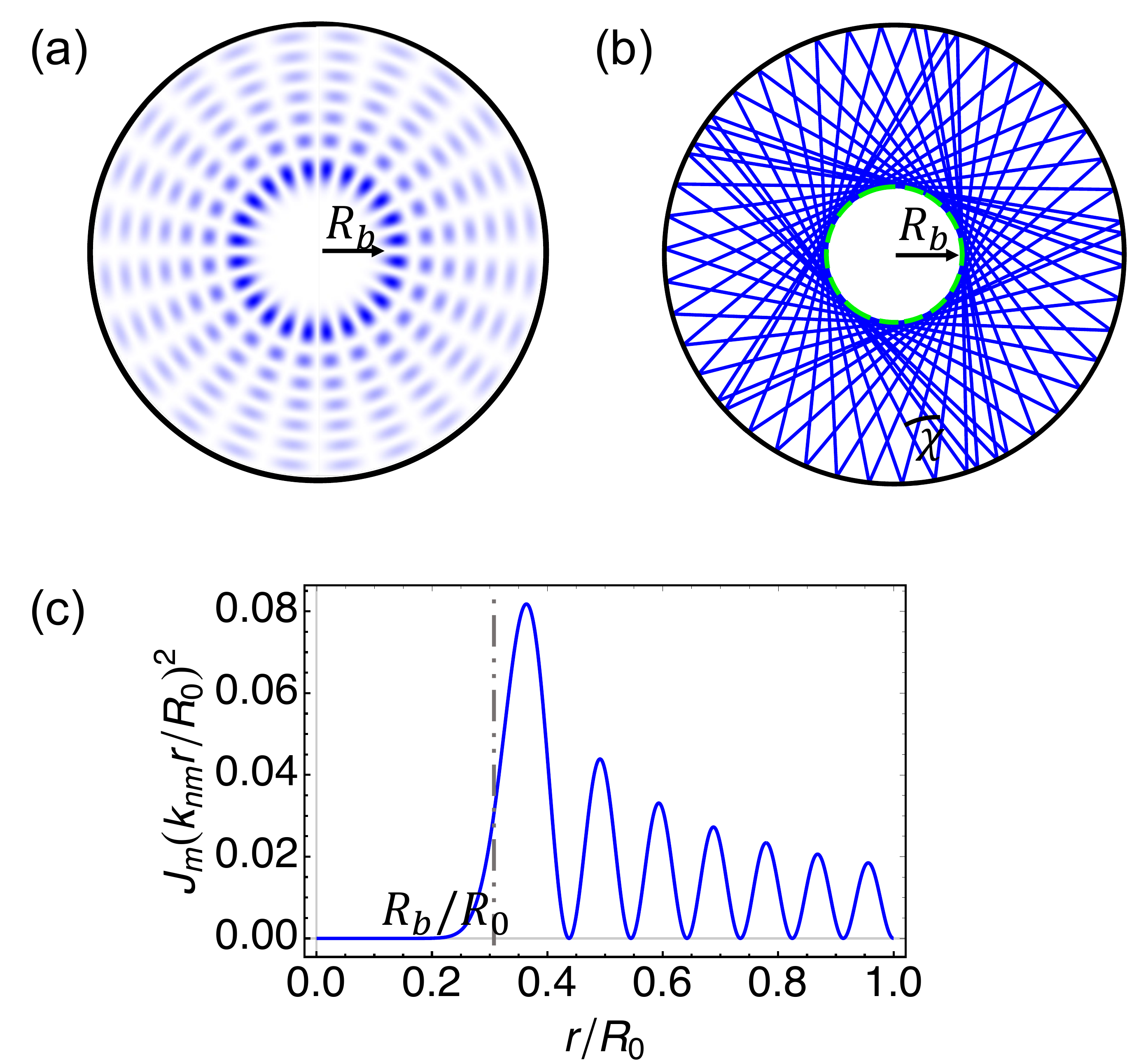}
    \caption{Circular billiard. (a) A typical quantum energy eigenfunction $|\psi_{nm}+\psi_{n,-m}|^2/2$; (b) a typical classical trajectory; (c) the square of a Bessel function $J_{m}(k_{nm}r)$ with $R_b$ indicated. In this figure, $n=7$ and $m=12$. }
    \label{fig:cbilliard}
\end{figure}

\subsection{Circular Billiard}
For a quantum particle of mass $M$ moving in a circular billiard of radius $R_0$, its energy eigenstates can be expressed analytically 
in terms of the Bessel function $J_m$,
\begin{equation}
    \psi_{nm}(r,\theta)=\frac{1}{\sqrt{\pi}R_{0}|J_{m}'(k_{nm}R_{0})|}J_{m}(k_{nm}r)e^{im\theta}~,
\end{equation}
where $n$ is the radial quantum number and $m$ is the angular quantum number with $k_{nm}$ being 
determined by the boundary condition
\begin{equation}
    J_{m}(k_{nm}R_0)= 0.
\end{equation}
It is clear that $ \ket{\psi_{nm}}$ and $\ket{\psi_{n,-m}}$ are degenerate with the same eigen-energy of $E_{nm}=\hbar^2k_{nm}^2/(2M)$. 
For simplicity, we choose the units in which $\hbar=M=1$ in the following discussion. 

A typical energy eigenfunction is plotted in Fig.\ref{fig:cbilliard}(a). An obvious feature is that the eigenfunction $\ket{\psi_{nm}}$ 
(or any linear superposition of $\ket{\psi_{nm}}$ and $\ket{\psi_{n,-m}}$)  is almost zero inside a circle of a certain radius $R_b$. 
The corresponding classical motion has a similar ``blank region".  As one can see clearly from Fig. \ref{fig:cbilliard}(b), 
for a classical particle bouncing elastically inside the circular billiard, if its initial angular momentum is $MvR_b$, 
it never moves inside the circle of radius $R_b$. Since the motion of a classical particle in a billiard is independent of the size
of its momentum $Mv$, the radius $R_b$ of such a blank region can be regarded as a kind of normalized angular momentum.  
With this understanding in mind,  for an energy eigenstate $\ket{\psi_{nm}}$, we define a normalized angular momentum ${\Re}_{nm}$ as 
\begin{equation}
    \Re_{nm}=\frac{\langle \psi_{nm}|L_{z}|\psi_{nm}\rangle}{\sqrt{2E_{nm}}}=\frac{m}{k_{nm}}. 
\end{equation}
As indicated in Fig.\ref{fig:cbilliard}(c), $\Re_{nm}$ defined in such a way can be regarded as  the radius of 
the ``blank region" of $\ket{\psi_{nm}}$. 

Our discussion above shows that the radius $ \Re_{nm}$ can be used to characterize the eigenstate $\ket{\psi_{nm}}$. To be precise, 
we introduce a classifier,
\begin{equation}
\label{eq:circ}
    C(R_b;|\psi_{nm}\rangle)=\left\{
    \begin{array}{rl}
    1 & \text{if } 0< \Re_{nm}<R_{b},\\
    0 & \text{if } R_{b}\leq  \Re_{nm}<R_{0}.
\end{array} \right.
\end{equation}
It says that if the radius $ \Re_{nm}$ of an eigenstate $\ket{\psi_{nm}}$ is smaller than $R_b$ then $C(R_b;|\psi_{nm}\rangle)=1$; 
otherwise it is zero. We consider an energy shell $\mathcal{S}(E_c, \Delta E) = [E_c-\Delta E/2, E_c+\Delta E/2]$, which 
is centered at $E_c$ and with a width of $\Delta E$. We are interested in how many  eigenstates in the shell 
$\mathcal{S}(E_c, \Delta E)$ have their blank region radii $\Re_{nm}<R_{b}$. 
For this purpose, we define a ratio 
\begin{equation}\label{eq:fRb_def}
 f(R_{b};E_{c},\Delta E)=\frac {\sum_{E_{nm}\in \mathcal{S}}C(R_{b};|\psi_{nm}\rangle)} 
 {{\rm number~of~eigenstates~in~} \mathcal{S}}\,.
\end{equation}
%where the nominator is effectively the total number of the eigenstates in the energy shell $\mathcal{S}(E_c, \Delta E)$.

\begin{figure}[ht]
    \centering
    \includegraphics[width=0.9\linewidth]{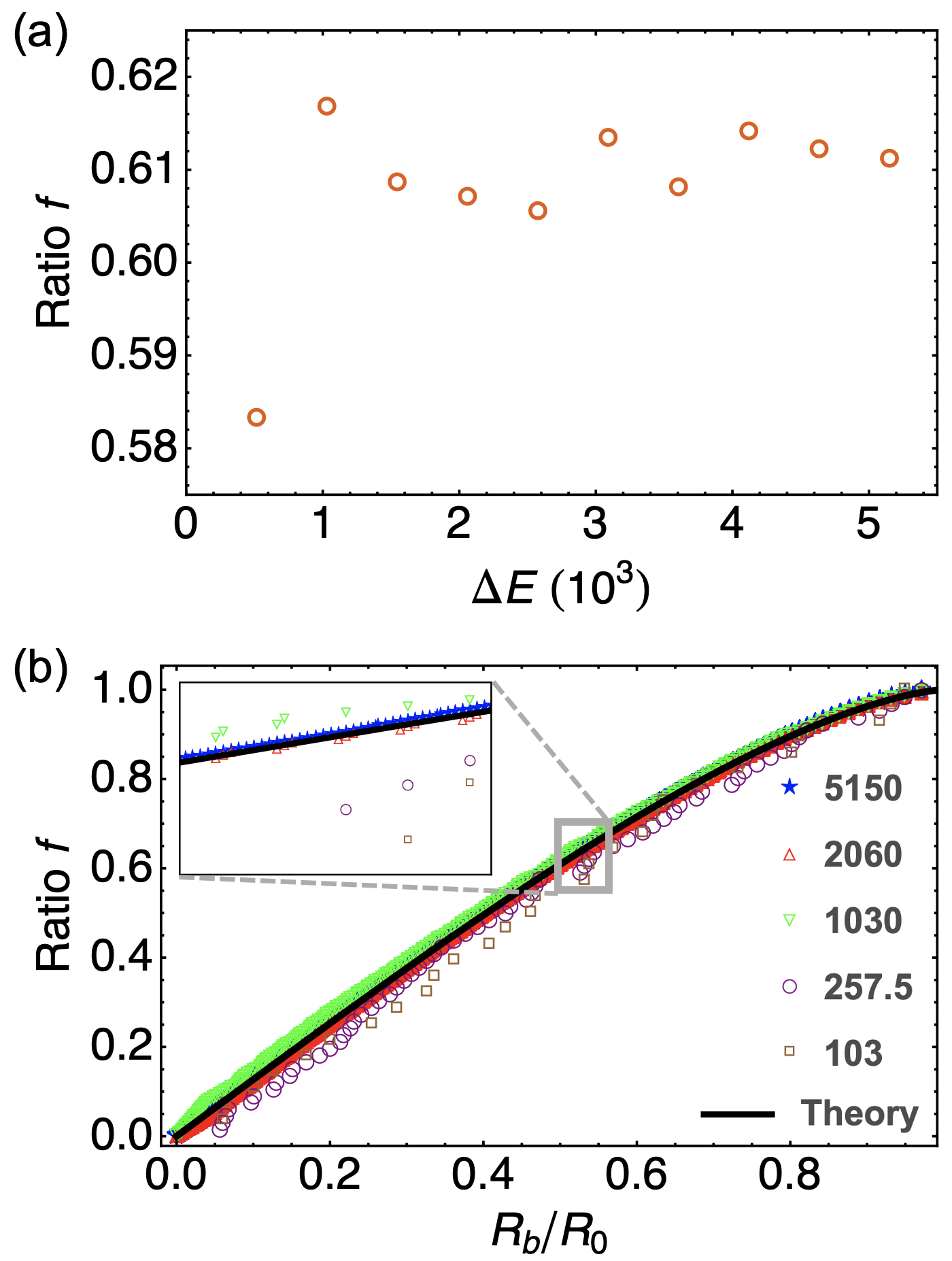}
    \caption{(a) Ratio $f(R_0/2;E_{c},\Delta E)$ as a function of $\Delta E$. 
    (b) Ratio $f(R_0/2;E_{c},\Delta E)$ as a function of $R_{b}/R$. 
    In this numerical results, we take $\hbar=1$ and fix $k_{c}=515$ (corresponding $E_{c}=k_{c}^{2}/2$), while changing $\Delta k$ from $0.1$ to $5$ (corresponding $\Delta E=2k_{c}\Delta k$). The narrowest energy window (i.e. $\Delta k=0.2$) has merely 28 energy eigenstates, thus $\Delta k$ cannot be further narrowed. Here we plot the ratio of energy eigenstates with ``blank region radius'' less then $R_{b}$ vs. $R_{b}$. Except for $\Delta k=0.1$ and $0.25$ (i.e. energy window is too narrow), other curve (with moderate or large energy window) embody a perfect self-similarity. The black line represents the theoretical result Eq.\ref{eq:fRb}.}
    \label{fig:fractal2}
\end{figure}

We have numerically computed the ratio $f(R_b;E_{c},\Delta E)$. One set of the results are plotted in Fig.\ref{fig:fractal2}(a), which shows
how the ratio $f(R_b;E_{c},\Delta E)$ changes with $\Delta E$ with $R_b$ fixed at $R_0/2$. It is clear from this figure 
that the ratio $f$ stays almost constant once the shell width $\Delta E$ is large enough. For this specific example, 
the figure shows that in a not-too-narrow energy shell,  there are alway about 61\% of the eigenfunctions $\ket{\psi_{nm}}$ 
whose blank region radius $R_b$ is smaller than $R_0/2$.  This is self-similarity. 
Fig.\ref{fig:fractal2}(b) shows that this kind of self-similarity exists for all values of $R_b$ not just for $R_b=R_0/2$. 
In particular, as seen from this figure, the curves $f(R_b;E_{c},\Delta E)$  
for different widths  $\Delta E$ approach a limiting curve when $\Delta E$  increases. 
Note that the above results are not sensitive to the center $E_c$ of the energy shell as long as it is not too close to the ground state.

\begin{figure}[!th]
    \centering
    \includegraphics[width=0.85\linewidth]{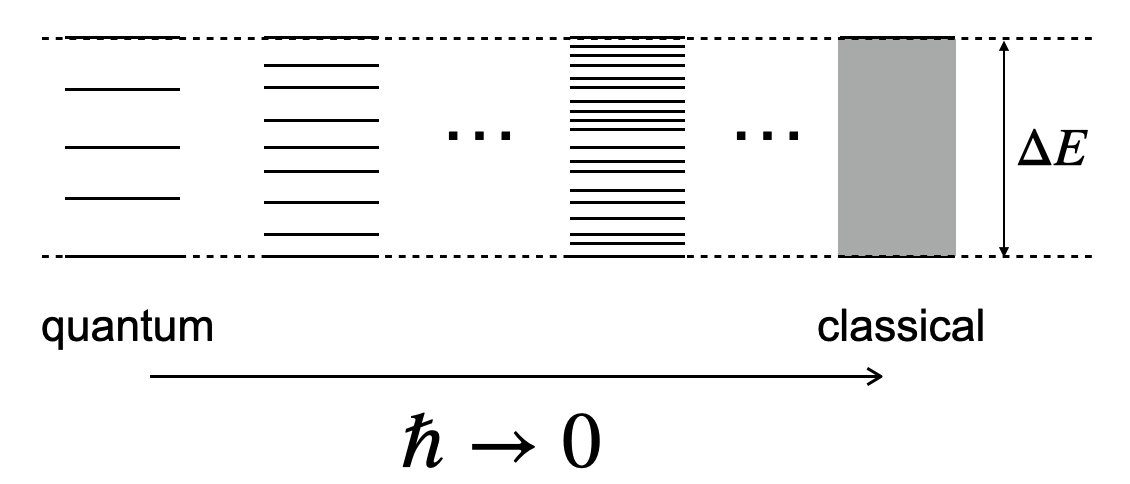}
    \caption{As the Planck constant $\hbar$ becomes smaller, more energy eigenstates enter a given energy shell. 
    In the  semi-classical limit $\hbar\rightarrow 0$, the energy spectrum becomes continuous.}
    \label{fig:hbarzero}
\end{figure}

\iffalse
For any quantum system,  when $\hbar$ becomes smaller,  
more energy eigenstates enter an energy shell $\mathcal{S}(E_c, \Delta E)$ with fixed center $E_c$ and 
width $\Delta E$ (see Fig. \ref{fig:hbarzero}).  For this billiard system, as its eigen-energy $E_{nm}=\hbar^2 k_{nm}/(2M)$, 
decreasing $\hbar$ by a factor of $w>1$ is equivalent to increasing the shell width $\Delta E$ by a factor of $w^2$.
In other words, the energy eigenstates now in the shell $\mathcal{S}(E_c, \Delta E)$
are effectively the energy eigenstates  in the bigger energy shell $\mathcal{S}(E_c, w^2\Delta E)$. 
Therefore, the self-similarity demonstrated in Fig.\ref{fig:fractal2} implies that at the smaller Planck constant $\hbar/w$, 
for any not-too-small energy shell within  $\mathcal{S}(E_c, \Delta E)$, there are a fraction of $f(R_b;\Delta E)$ eigenstates 
whose blank region radius is smaller than $R_b$. 
%Here $E_c$ can be any value as long as the smaller shell lies within the larger shell. 
Since the system becomes classical in the limit of $\hbar\rightarrow 0$, the ratio $f(R_b;\Delta E)$ is likely to have 
a classical interpretation. This is indeed the case as we shall see. 
\fi

For any quantum system,  when $\hbar$ becomes smaller,
more energy eigenstates enter an energy shell $\mathcal{S}(E_c, \Delta E)$ with fixed center $E_c$ and 
width $\Delta E$ (see Fig. \ref{fig:hbarzero}). For this billiard system, as its eigen-energy $E_{nm}=\hbar^2 k_{nm}/(2M)$, 
decreasing $\hbar$ reduces the gap between nearest energy levels and is roughly equivalent to enlarge the width of energy shell.
(See appendix \ref{app_deltaE} for a detail discussion of the relation between decreasing $\hbar$ and increasing $\Delta E$.)
Therefore, the self-similarity demonstrated in Fig.\ref{fig:fractal2} implies that at the smaller Planck constant, the fraction of $f(R_b;\Delta E)$ would get little changed.
%Here $E_c$ can be any value as long as the smaller shell lies within the larger shell. 
Since the system becomes classical in the limit of $\hbar\rightarrow 0$, the limit of the ratio $f(R_b;\Delta E)$ is likely to have 
a classical interpretation. This is indeed the case as we shall see.

\begin{figure}[ht]
    \centering
    \includegraphics[width=0.5\linewidth]{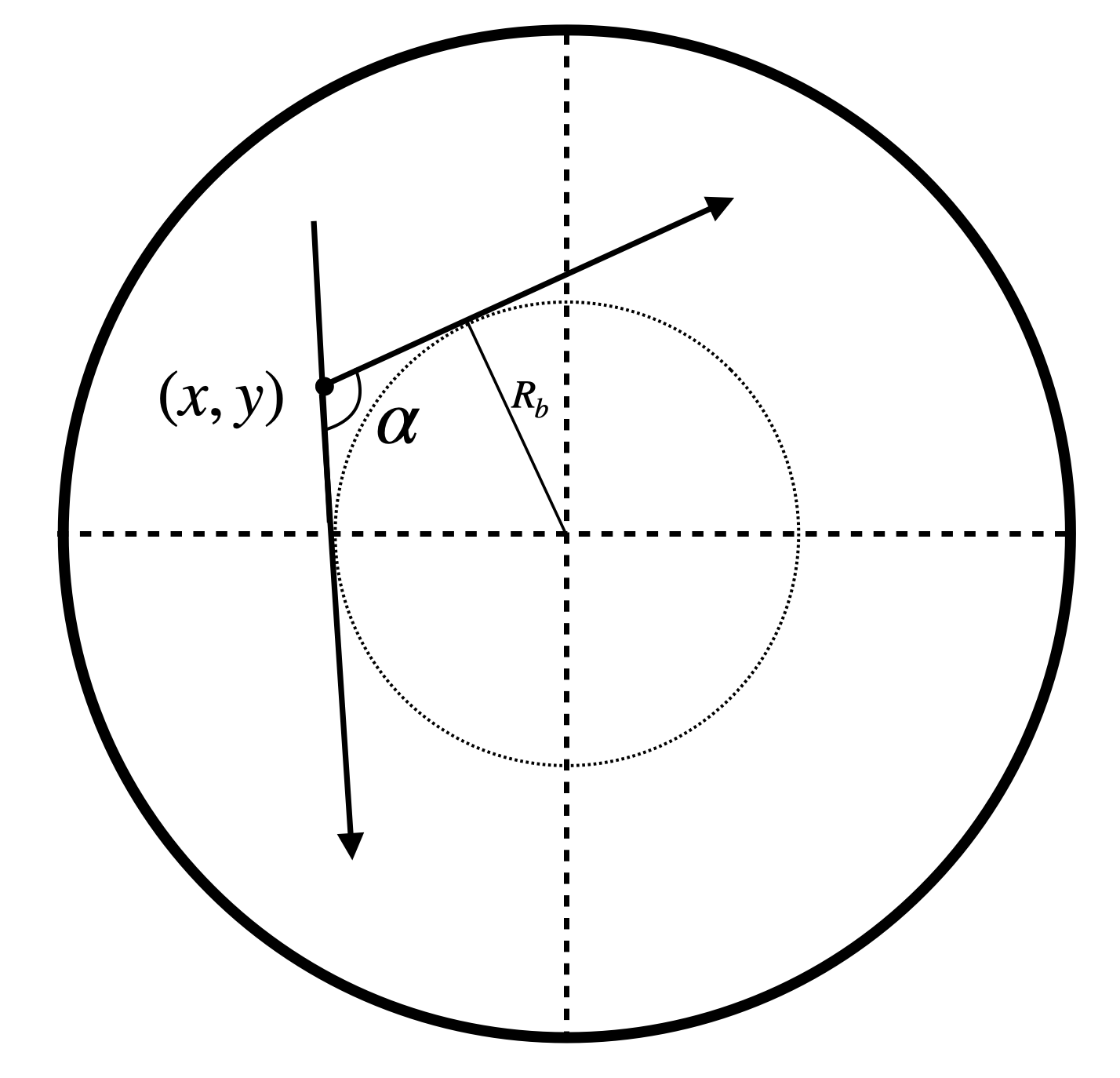}
    \caption{In a classical circular billiard, for a particle starting at $\mathbf{r}=(x,y)$, only when its direction of motion is limited inside
    the angle of $\alpha$, it can enter a circle of radius of $R_b$. Otherwise, it always stays outside of the circle.}
    \label{fig:Rb}
\end{figure}

Let us consider the classical circular billiard. For a classical system, the energy shell $\mathcal{S}(E_c, \Delta E)$ specifies
a volume in its phase space. For a billiard system, as its dynamics is the same for all different energies, we 
can focus on an isoenergetic surface in the volume. We define another ratio 
\begin{equation}
    g(R_{b}; E)=\frac{\int_{<R_b}dx dy dp_{x} dp_{y}}{\int_{T} dx dy dp_{x} dp_{y}}\,.
\end{equation}
Here the nominator is the phase-space volume of a constant energy surface with energy $E=p^{2}/2$. For the trajectories in a classical billiard 
are the same for different momentum, we take $p=1$ for simplicity. The denominator is the  volume occupied by all the trajectories with 
blank region radius smaller than $R_b$.  For a trajectory starting at  point $\mathbf{r}=(x,y)$ in the circular billiard, it is completely determined by
the direction of its momentum $\mathbf{p}$.  For this trajectory  to enter the circle of radius $R_b$, the direction of its motion
must be limited in the angle $\alpha(x,y;R_{b})=2\arcsin(R_{b}/r)$, where $r=\sqrt{x^{2}+y^{2}}$, as shown in Fig.\ref{fig:Rb}. 
As a result, we have 
\begin{equation}
    g(R_{b})=\frac{\int_{r<R_{b}}  \alpha(x,y;R_{b})dxdy}{\pi\int_{0<r<R_{0}} dxdy}.
\end{equation}
Evaluating this integral in the polar coordinate, we obtain
\begin{equation}\label{eq:fRb}
    g(R_{b})=\frac{2}{\pi}(u\sqrt{1-u^{2}}+\arcsin u),
\end{equation}
where $u=R_{b}/R_{0}$. This theoretical result is plotted as a black line  in Fig.~\ref{fig:fractal2}(b), 
where we see that it  agrees very well with our numerical results for $f(R_b;E_{c},\Delta E)$.

\begin{figure}[h]
    \centering
    \includegraphics[width=0.6\linewidth]{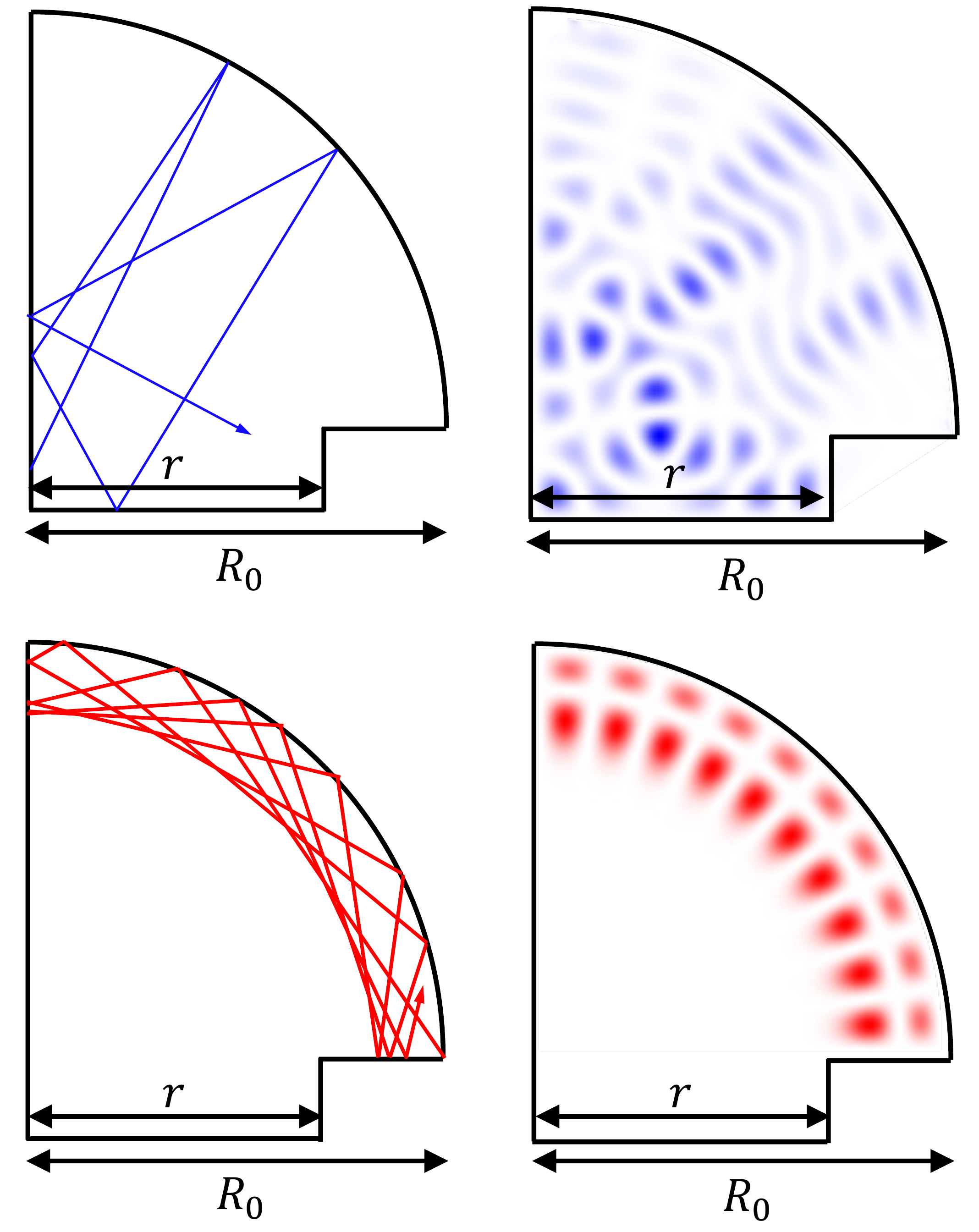}
    \caption{Mushroom billiard which is made of a quarter of circle and a rectangle. For the motion with the blank
    region radius $R_b$ smaller than $r$, the horizontal length of the rectangle, it is chaotic; otherwise, it is integrable.  }
    \label{fig:mushroom}
\end{figure}

The circular billiard is an integrable system, but our results can be safely generalized for non-integrable or chaotic systems. 
The mushroom billiard in Fig. \ref{fig:mushroom}, which is obtained by adding a rectangular stalk to a quarter of circle, 
is a non-integrable system, which has both integrable motions and chaotic motions\cite{mushroom}. When 
the blank region radius or normalized angular momentum $R_b$ is bigger than $r$, the motion is integrable; otherwise, it is chaotic. 
We find that the number of chaotic energy eigenstates, e.g., the one in the upper-right corner of Fig. \ref{fig:mushroom}, in an energy shell
is proportional to the volume of chaotic trajectories in the classical phase space. For this system, $R_b$ is an indicator of chaotic motion.

\subsection{General discussion}
With the circular  billiard, we have found a self-similarity in energy eigenstates that is intimately related to  
the classical dynamics. This  is in fact  a general feature that  exists in all quantum systems 
as our analysis below shows. 

\begin{figure}[ht]
    \centering
    \includegraphics[width=0.85\linewidth]{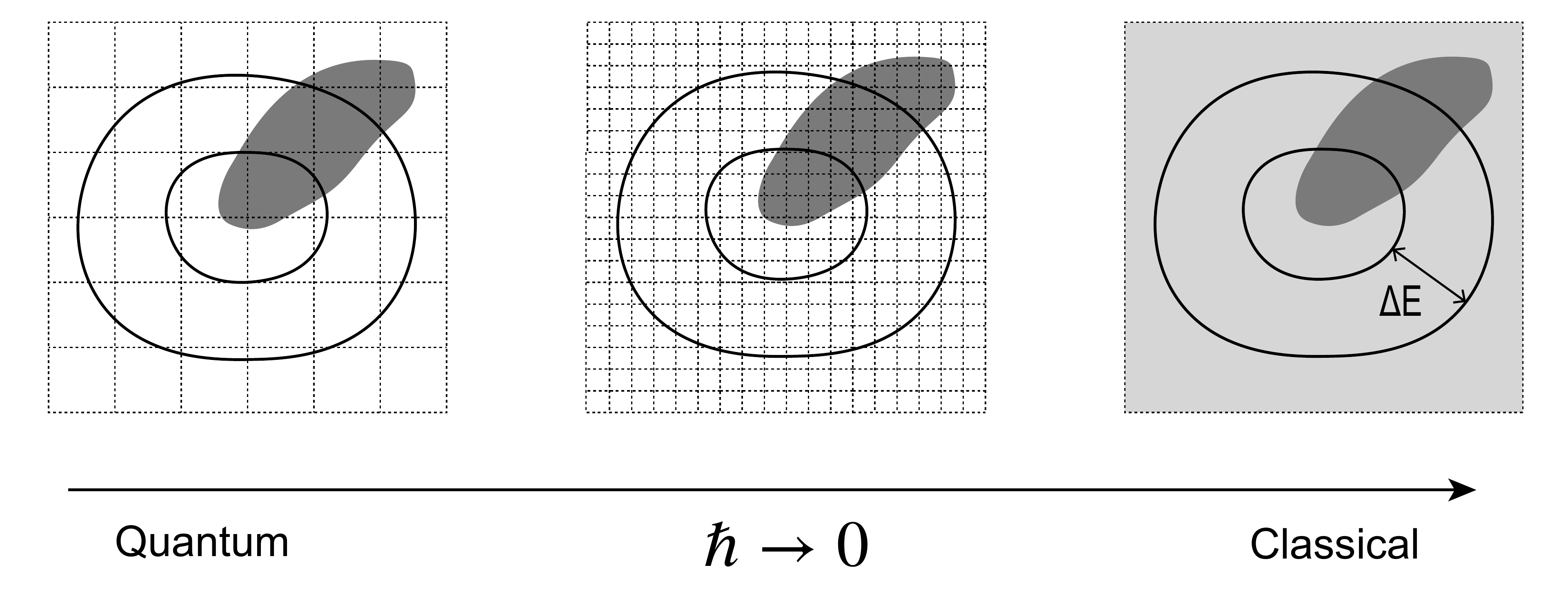}
    \caption{The schematic of  a quantum system approaching its semiclassical limit $\hbar\rightarrow 0$ in phase space. 
    The two black curves  represent two constant energy surfaces with difference $\Delta E$; the area enclosed by 
    them are the volume of a given energy shell ${\mathcal S}(E_c,\Delta E)$.  The dark shaded area is
     a region where the system has a certain physical property $A$.  As $\hbar$ decreases towards zero,  
     the Planck cells (squares in the figure) become smaller. As a result, there are more energy eigenstates in
     the energy shell ${\mathcal S}(E_c,\Delta E)$ and more eigenstates have property $A$. However, 
     the proportional of energy eigenstates having property $A$ in the energy shell ${\mathcal S}(E_c,\Delta E)$ quickly approach
     a limit, which is the overlap of the energy shell and the dark area. 
     }
    \label{fig:pspace}
\end{figure}

We consider a general quantum system, which has a well-defined classical counterpart. Its quantum 
phase space can be obtained by dividing the classical phase space into Planck cells\cite{otoc} 
as shown in Fig.\ref{fig:pspace}.  An energy shell ${\mathcal S}(E_c,\Delta E)$ 
in phase space is a volume enclosed by two constant energy surfaces around $E_c$, which are plotted as 
two black curves in Fig.\ref{fig:pspace}. The dark shaded area in the figure is a collection of all the states (quantum or classical)  
that have physical property $A$. For the above example of the circular billiard, the property $A$ is the blank region 
radius smaller than $R_b$.  According to quantum mechanics, the number of Planck cells in a phase space volume
is the same as the number of energy eigenstates\cite{pathria}. As a result, the number of energy eigenstates with property $A$ 
in the energy shell ${\mathcal S}(E_c,\Delta E)$ is equal to the number of Planck cells in the overlap region, 
the dark area between two black curves in Fig.\ref{fig:pspace}.  When $\hbar$ decreases, the Planck cells become smaller 
so that the number of energy eigenstates in different categories, e.g., in an energy shell or having property $A$, increases. 
However, the ratio between numbers in these different categories will quickly saturate and reach a limit that is set by the 
ratio between volumes of corresponding categories in the classical phase space. 
For the circular billiard, the former is $f(R_b;E_{c},\Delta E)$ and the latter is $g(R_b;E_c)$.

The above analysis motivates us to generalize the classifier $C$ and  
related ratio $f$ in Eqs.(\ref{eq:circ},\ref{eq:fRb_def}).  
For a given quantum system, a classifier $C$ maps its energy eigenstates 
into binary value $C(\ket{E}) \in \{0, 1\}$, that is, divides the eigenstates  into two groups. Formally, 
we can write
\begin{equation}
    C(\xi;\ket{E}) = \begin{cases}
    0 & \mathcal{A}(\ket{E}) < \xi \\
    1 & \mathcal{A}(\ket{E}) \geq \xi
    \end{cases},
\end{equation}
where $\mathcal{A}$ is a function that characterizes  physical property $A$ of an eigenfunction $\ket{E}$ 
and $\xi$ is  a certain value.  The choice of property $A$  and related function $\mathcal{A}$ depends on the quantum system
that is being considered. For the circular billiard, we have chosen the blank region radius $R_b$ and related function $\Re_{nm}$. 
For a quantum chaotic system, one possible choice is 
the effective occupation $\mathcal{A}(\ket{E}) = (\sum_{x} |\braket{x|E}|^4)^{-1}$\cite{qkam}, which 
characterizes how widely  a wave function spreads in space. 
More examples will be given later.  Note that one can certainly define a classifier that maps the 
energy eigenstates into three or more groups. We for simplicity focus on the above definition. 
%We emphasize that if the quantum system has a well-defined classical limit, 
%the property $A$ must be meaningful for the corresponding classical system. 

For an energy shell $\mathcal{S}(E_c, \Delta E, \hbar) = \{\ket{E}: E\in [E_c-\Delta E/2, E_c+\Delta E/2]\}$, 
with the above classifier, we define the following ratio
\begin{equation}\label{eq:ffunction}
    f(\wp,E_{c},\Delta E) = \frac {\sum_{\ket{E} \in \mathcal{S}}C(\ket{E})}
     {{\rm number~of~eigenstates~in~} \mathcal{S}}.
\end{equation}
Here $\wp$ is a parameter that controls the number of energy eigenstates in the energy shell $\mathcal{S}(E_c, \Delta E)$. 
When the quantum system has a well-defined classical limit, $\wp$ is just the Planck constant $\hbar$
or the effective Planck constant. 
The self-similarity in energy eigenstates means that the ratio $f(\wp,E_{c},\Delta E) $ is independent of 
the control parameter $\wp$ and the width $\Delta E$ as long as the number of eigenstates in the shell $\mathcal{S}$ is
statistically large enough.  In particular, one can divide the shell $\mathcal{S}$ into many small sub-shells and tune 
the control parameter $\wp$ so that each sub-shell contains enough energy eigenstates. In this case, within statistical fluctuations, 
the ratio $f$ is the same for every sub-shell. 

\iffalse
The ratio $f(\wp,E_{c},\Delta E)$ has a very weak dependence on $E_c$, the center of the energy shell. 
By weak dependence, we mean that $f(\wp,E_{c},\Delta E)$ varies with $E_c$ only over a very large energy scale. 
Consider water as an example. The physical property of water with $E_c/k_B$  around the boiling temperature
is very different from its property with $E_c/k_B$  around the freezing temperature. However, for a quite large energy range 
between the freezing and boiling temperatures, water is just a liquid and its property does not change much. 
For systems similar to the circular billiard, $E_c$ does not affect significantly the system as long as it is not too close to the lowest quantum
energy. In this work, we focus on the situations where the dependence of $f$ on $E_c$ can be safely neglected. \fi

For a quantum system with a well-defined classical counterpart, such self-similarity is rooted in 
the correspondence between the energy eigenstates and invariant distributions in classical phase space\cite{Berry_1977,knauf}. 
Without loss of generality, we choose $\wp=\hbar$ and focus on time-independent systems in the following discussion. 
By a well-defined classical counterpart, we mean that the quantum dynamics starting with a wave function that 
is well-localized in phase space follows the classical trajectory in the semiclassical limit $\hbar\rightarrow 0$. 
In  the Planck cell notation\cite{otoc}, such a correspondence can be written as
\begin{equation}
    \lim_{\hbar \rightarrow 0} |\braket{x' | \hat U(t) | x}|^2 = \delta_{x', g_t x},
\end{equation}
in which $\hat U(t) = e^{- i \hat H t/\hbar}$ is the propagator of time evolution during $t$ while $g_t$ is the corresponding classical time evolution by canonical equations\cite{Arnold}. The basis $\ket{x} = \ket{Q, P}$ is the Planck cell basis at a discretized phase space\cite{otoc}. 
As a result, for an energy eigenstate $\ket{E}$, we have 
\begin{equation}
    \lim_{\hbar \rightarrow 0} |\braket{x|E}|^2 = \lim_{\hbar\rightarrow 0}|\bra{x} \hat U(t) \ket{E}|^2 = \lim_{\hbar\rightarrow 0}|\braket{g_t^{-1} x|E}|^2.
\end{equation}
This shows that at the semiclassical limit, each energy eigenstates becomes a distribution in phase space which is invariant 
under classical dynamics\cite{Berry_1977,knauf}. 

For a classical system, its isoenergetic surface is usually filled with different invariant distributions, which do not overlap with each other\cite{phys_dis}. 
The \poincare sections in Figs.\ref{fig:top_psect1}\&\ref{fig:clkr} offers some glimpses of such a structure: the invariant 
distributions represented by the chaotic seas do not overlap with the distributions represented by smooth lines in integral islands.
Consider an energy shell ${\mathcal S}$ with a very small width $\Delta E$ so that each isoenergetic surface within the shell
is filled with similar non-overlapping invariant distributions. In this way, with $\Delta E$ one can legitimately say that
an invariant distribution occupies a volume in phase space. 
Due to the quantum-classical correspondence discussed above, 
these different invariant distributions are the limits of different energy eigenstates when $\hbar$ goes to zero. 
As the energy is continuous in classical mechanics, the energy shell width $\Delta E$ can be arbitrary small. 
And for a given width $\Delta E$, no matter how small it is, we can always choose a small enough $\hbar$ so that 
there are large number of eigenstates in the shell ${\mathcal S}$, in which the number of each type of eigenstates 
is proportional to the volume of the corresponding invariant distribution. This gives rise to the self-similarity 
that we have found in eigenstates. 

Our above analysis has been done with quantum systems that have well-defined classical limits. However, 
such self-similarity appears very general and exists in all quantum systems. This is indicated by 
our numerical computation in the next section, where  a model of 
Heisenberg spin chain is studied. This system has no well-defined classical limit, and we still find self-similarity 
in its eigenstates. It is not clear why self-similarity exists in such quantum systems. 

Before we present more examples, we use an analogy to summarize our finding. For a quantum system, 
if we regard each of its energy eigenstates as a small ball, then all the eigenstates lie on a one-dimensional line 
in the order of their corresponding eigen-energies. Such a line has at least one end, which is the ground state. 
Suppose that a fraction $f$ of these balls are red, representing that the corresponding eigenstates have property $A$. 
We find that these red balls are thoroughly mixed with other balls. As a result, for any segment of line that contains 
large number of balls, the fraction of red balls on this segment is exactly $f$ if we ignore the statistical fluctuations. 

In quantum chaotic systems with no apparent symmetries, degeneracy rarely happens. People often refer to it as energy level repulsion. 
Our finding can also be regarded as a repulsion phenomenon: energy eigenstates with similarity properties tend to ``repel" each other 
and scatter rather evenly among other eigenstates.  And this kind of repulsion exists for all quantum systems, not limited to chaotic systems. 

%%%%%%%%%%%%%%%%%%%%
\section{Examples exhibiting Self-similarity}\label{sec:numerical}
Below are three examples. The first example, quantum coupled top, is a time-independent system; 
the second example, quantum kicked rotor, is a periodically-driven system; the third example, a Heisenberg chain, is a quantum 
system that has no well defined classical limit.  Self-similarity in energy eigenstates is evident in all of them. 
The third examples suggests that the self-similarity exists also in quantum systems that have no well defined classical limits. 

\subsection{Quantum Coupled Top}
The quantum coupled top is a famous model that were used to study quantum chaos\cite{Peres,Feingold}. It describes the interaction 
between two identical angular momentum 
$\hat{\boldsymbol{L}}_{1}$ and $\hat{\boldsymbol{L}}_{2}$, which is governed by the following Hamiltonian\cite{doubletop,doubletop2}
\begin{equation}\label{eq:doubletopH}
    \hat{H}=\hat{L}_{1z}+\hat{L}_{2z}+\frac{\mu}{J}\hat{L}_{1x}\hat{L}_{2x},
\end{equation}
where $J$ is the magnitude of the angular momentum, and $\mu$ denotes the coupling constant. 
%\red{We take $\hbar=1$, and $1/J$ as the effective Planck constant, i.e. $h_{\mathrm{eff}}=1/J$, so the magnitude of matrix elements for angular momentum will not change with $h_{\mathrm{eff}}$. The system tend to classical as $J\rightarrow+\infty$.}

\begin{figure}[hbtp!]
    \centering
    \includegraphics[width=1\linewidth]{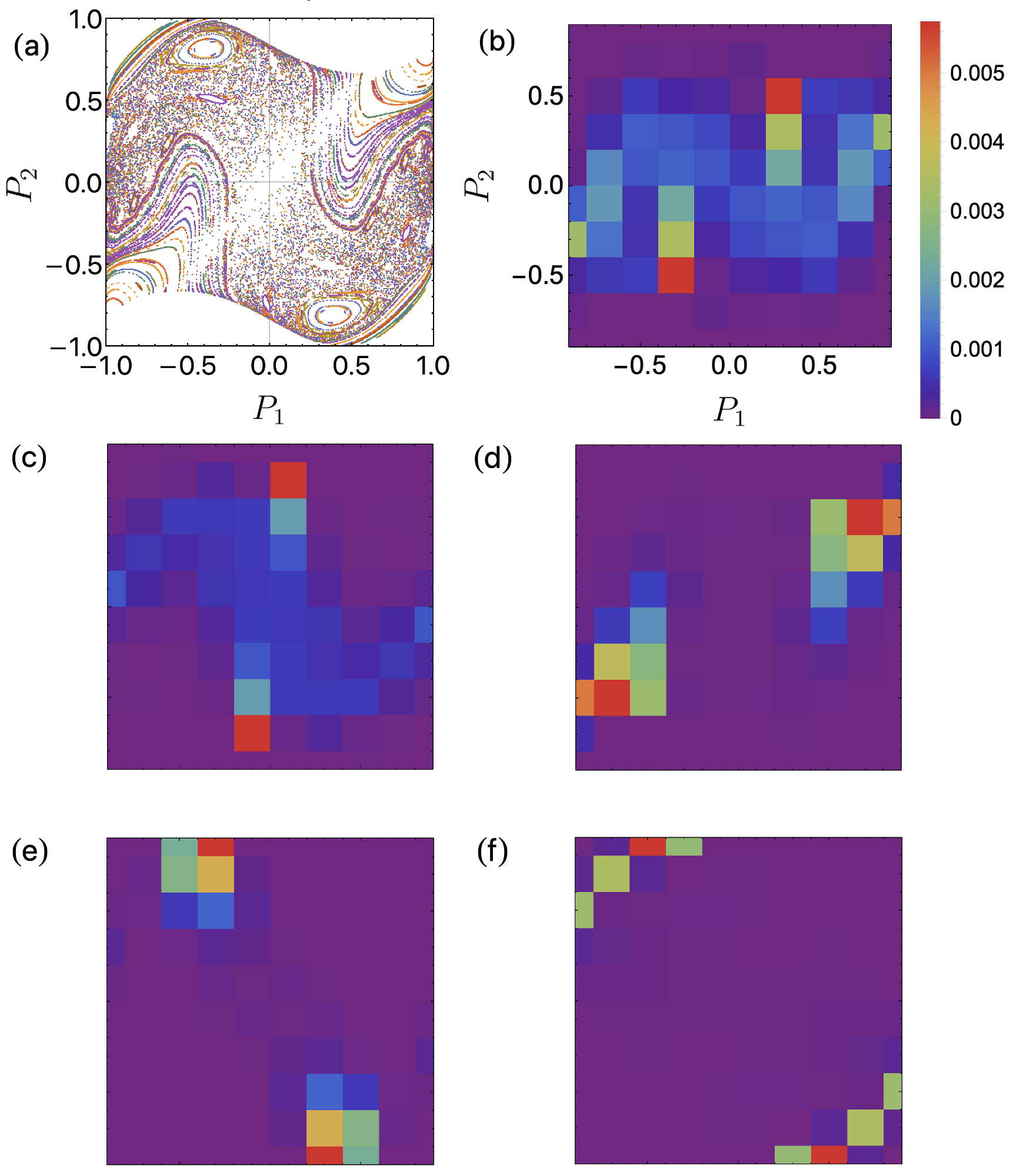}
    \caption{(a) \poincare sections for the  classical coupled top model with energy $E_{c}=-0.9$.
    (b,c,d,e,f) are the 3194 , 3196, 3198, 3202, 3207th energy eigenstates (energy eigenvalues are $-0.9018, -0.9007, -0.8998, -0.8970, -0.8939$), respectively, 
    in the quantum phase space. The parameters are $\mu=0.5$ and $Q_{1}=\pi$. }
    \label{fig:top_psect1}
\end{figure}

This system has a well defined classical counterpart, whose Hamiltonian is obtained by 
simply replacing the operators $\hat{\boldsymbol{L}}_{1}$ and $\hat{\boldsymbol{L}}_{2}$ with 
two variables of angular momentum $\boldsymbol{L}_{1}$ and $\boldsymbol{L}_{2}$. 
For the classical model, it is convenient to introduce a different set of canonical variables 
\begin{equation}\label{eq:L}
    \textstyle{\boldsymbol{L}_{i}=J(\sqrt{1-P_{i}^{2}} \cos Q_{i},\sqrt{1-P_{i}^{2}}\sin Q_{i}, P_{i})},
\end{equation}
where $i$ is either 1 or 2. The classical Hamiltonian becomes
\begin{equation}
    \textstyle{H=J(\sqrt{1-P_{1}^{2}}\cos Q_{1}+\sqrt{1-P_{2}^{2}}\cos Q_{2}+\mu P_{1}P_{2}).}
\end{equation}
(rigorous derivation of quantum-classical correspondence is illustrated in appendix \ref{app_pathintegral})
The classical dynamics is described by the following canonical equations:
\begin{equation}\label{eq:doubletop_eom}
    \textstyle{\dot{P_{i}}=-\sqrt{1-P_{i}^{2}}\sin Q_{i}}~,~
    \dot{Q_{i}}=\frac{P_{i}}{\sqrt{1-P_{i}^{2}}} \cos Q_{i}+\mu P_{j},
\end{equation}
where $i,j\in\{1,2\}$ and $i\not=j$.  The classical dynamics is a mixture of 
regular and chaotic motions as indicated by the classical \poincare section at $Q_{1}=\pi$ with 
$\sin Q_{2}>0$ in Fig.~\ref{fig:top_psect1}(a).  The parameters for the figure are 
$J=2,~\mu=0.5$ and the energy $E_{c}=-0.9$.

To plot energy eigenstates in phase space, we first construct quantum phase space by 
dividing phase space into $L^{4}$ Plank cells, i.e. $Q_{i}=0,\frac{2\pi}{L},\dots,\frac{2\pi}{L}(L-1)$,~and~$P_{i}=\frac{-mL}{j},\frac{-(m-1)L}{j},\dots,\frac{mL}{j}$. Here $L=2m+1$ and $L^{2}=2j+1$. Limited by computational resources, we take $L=10$. 
For each Plank cell, we assign a localized quantum state\cite{qkam,otoc,phys_dis} 
\begin{equation}
    \vert Q_{1},P_{1};Q_{2},P_{2}\rangle=\vert Q_{1},P_{1}\rangle\otimes\vert Q_{2},P_{2}\rangle,
\end{equation}
where $\vert Q,P\rangle$ is defined as\cite{otoc}
\begin{equation}\label{eq:ketQP}
    \vert Q,P\rangle_{j}=\frac{1}{\sqrt{L}}\sum_{n=-m}^{m} e^{-iQ(n+Pj)} \vert n+Pj\rangle.
\end{equation}
The subscript $j$ meas that Planck cells $\vert Q,P\rangle$ are dependent on angular momentum quantum number $j$. We will neglect $j$ afterwards for simplicity.
These quantum states not only form a complete basis for the system but also 
have excellent classical meaning: the average angular momentum of state $\vert Q,P\rangle$ is just the same as Eq. \ref{eq:L} and
the uncertainties of $J_{x},J_{y},J_{z}$ are all proportional to $\hbar\sqrt{j}$, which decreases as $1/\sqrt{\hbar}$ with 
decreasing $\hbar$ and fixed $J$, see appendix \ref{app_planckcell}.

As  it is impossible to plot a complete energy eigenstate in the four dimensional phase space, we plot a section, 
which we call quantum \poincare section. In our calculation, for an energy eigenstate $|E\rangle$, 
we first set the section  at $Q_{1}=\pi$ and then compute the projection amplitude, 
$\rho_{Q,\mathrm{sec}}(P_{1},P_{2};E)=|\langle \pi,P_{1};Q_{2},P_{2}|E\rangle|^2$, where
\begin{equation}\label{eq:Q2}
    Q_{2}=\arccos\left[\frac{(H/J-\mu P_{1}P_{2}-\sqrt{1-P_{1}^{2}}\cos Q_{1})}{\sqrt{1-P_{2}^{2}}}\right].
\end{equation}
The results  $\rho_{Q,\mathrm{sec}}$ for five different eigenstates are shown in Figs.~\ref{fig:top_psect1}(b-f). 
The energy eigenvalues are chosen around  $E=-0.9$, the energy for the classical \poincare section 
in Fig.~\ref{fig:top_psect1}(a).  It is clear that each quantum \poincare section resembles a part of  the classical \poincare section. 
For example,  Fig.~\ref{fig:top_psect1}(d) resembles integrable islands located at the upper right and lower left 
corners of Fig.~\ref{fig:top_psect1}(a); Fig.~\ref{fig:top_psect1}(c) corresponds to the whole chaotic sea in Fig.~\ref{fig:top_psect1}(a); Fig.~\ref{fig:top_psect1}(e) resembles two integrable island located at the upper right and lower left corners 
of Fig.~\ref{fig:top_psect1}(a). The most interesting is that the five quantum \poincare sections combined 
just fill up the classical \poincare section.  This feature is general in our numerical results: 
for any classical \poincare section at a given energy $E_c$, we can always find energy eigenstates around $E_c$ whose 
quantum \poincare sections just fill up the classical \poincare section. This feature is a signature of self-similarity in energy eigenstates.

\begin{figure}[h]
    \centering
    \includegraphics[width=0.99\linewidth]{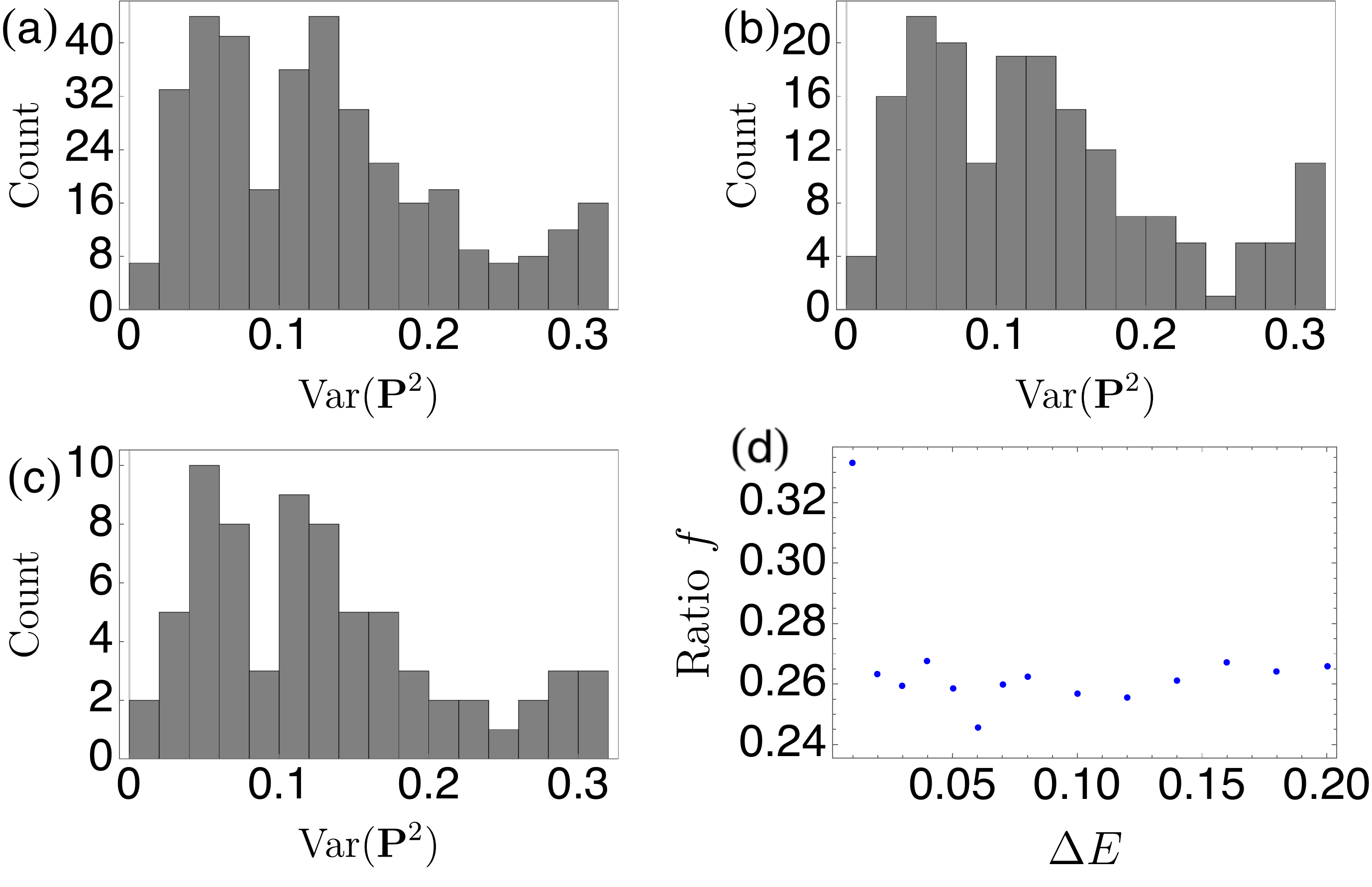}
    \caption{Histograms the variance $\mathrm{Var}(\mathbf{P}^{2})$ at three different widths  
    (a) $\Delta E=0.2$, (b)$\Delta E=0.1$, (c)$\Delta E=0.04$. (d) The ratio function $f(E_c,\Delta E)$ with the threshold $\delta=0.16$. 
    The energy shells are all centered at $E_{c}=-0.9$. }
    \label{fig:top_classifier}
\end{figure}

We next put the above observation on a quantitative ground by focusing on how wide   
the eigenstates spread in phase space. For an energy eigenstate $\ket{E}$ and its  distribution $\rho_{Q,\mathrm{sec}}$, we define
the following variance 
\begin{equation}
    \operatorname{Var}(\mathbf{P}^{2})\equiv\frac{\sum_{P_{1}}\sum_{P_{2}\geq0} \rho_{Q,\mathrm{sec}}(\mathbf{P})(\mathbf{P}-\bar{\mathbf{P}})^{2}}{\sum_{P_{1}}\sum_{P_{2}\geq0} \rho_{Q,\mathrm{sec}}(\mathbf{P})},
\end{equation}
where we abbreviate $(P_{1},P_{2})$ as a vector $\mathbf{P}$ and denote $\bar{\mathbf{P}}$ as the mean of $\mathbf{P}$ for the left half of \poincare section.  If $\ket{E}$ is chaotic, its $\operatorname{Var}(\mathbf{P}^{2})$ is large,  while $\operatorname{Var}(\mathbf{P}^{2})$ will be small if it is regular. 
The classifier is defined as
\begin{equation}
    C(\delta;\ket{E})=\left\{
    \begin{array}{rl}
    0 & \text{if } 0<\operatorname{Var}(\mathbf{P}^{2})<\delta\\
    1 & \text{if } \operatorname{Var}(\mathbf{P}^{2})\geq\delta
    \end{array} \right. ,
\end{equation}
where $\delta$ is a chosen threshold. For this classifier, the control parameter is $\Delta E$, the width of the energy shell. 
Our numerical results are summarized in Fig.\ref{fig:top_classifier}. The first three panels (a,b,c) are  the histograms of the variance 
for different widths $\Delta E$. They are very similar to each other. In the last panel (d),  
the ratio function $f(E_{c},\Delta E)$ in Eq.~\ref{eq:ffunction} is plotted, and it saturates quickly with $\Delta E$. 
They all demonstrate the self-similarity in eigenstates.

For this model, the last numerical check is done by counting the numbers of integrable and chaotic eigenstates. 
We compute all the eigenstates in the energy shell $[-0.95,-0.85]$, then examine each of them to see whether it is 
an integrable or chaotic eigenstate. We find that there are $N_{I}=122$ integrable eigenstates and $N_{C}=59$ chaotic ones. 
The volume of integrable islands and the  total volume of the constant energy surface at $E_c=-0.9$ in classical phase space 
are evaluated numerically. Our result is $V_{I}/V=0.60\pm 0.02$,  close to $N_{I}/(N_I+N_C)=0.68$, 
confirming the self-similarity. For details, see appendix \ref{app_volume}.

\subsection{Quantum Kicked Rotor}\label{subsec_kr}

\begin{figure}
    \centering
    \includegraphics[width=0.8\linewidth]{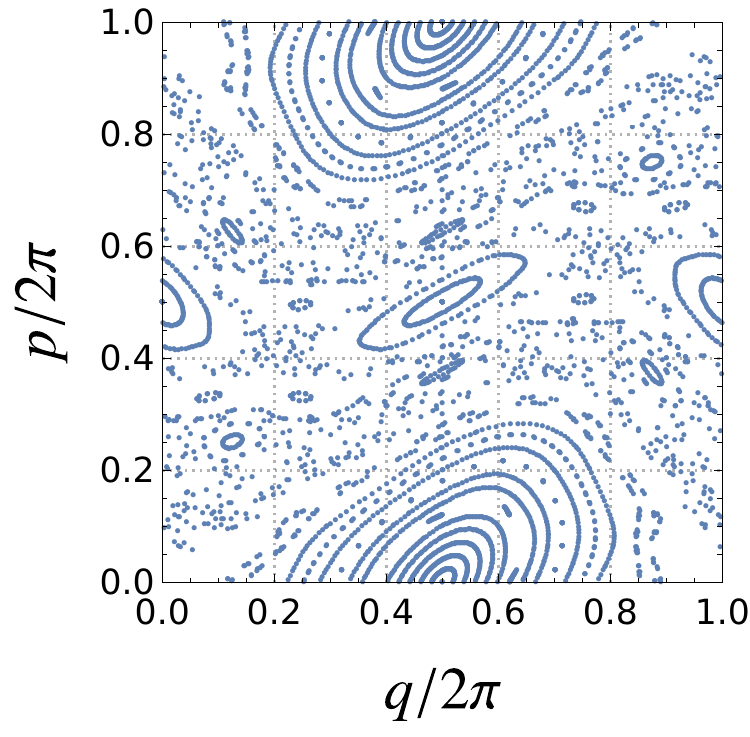}
    \caption{The \poincare section of the classical kicked rotor. The initial points are uniformly sampled along $p$-axis with fixed $q$ near $\pi$. }
    \label{fig:clkr}
\end{figure}

The Hamiltonian of kicked rotor is~\cite{Chirikov:2008}
\begin{equation}
    H = \frac {p^2} 2 + K \cos q \sum_{n=-\infty}^{\infty} \delta(t - n)\,,
\end{equation}
where $K$ is the kicking strength. Its classical dynamics can be reduced to the Chirikov map on the toric phase 
space $(q, p) \in [0, 2\pi)^{\otimes 2}$ by focusing the state before each kick, 
i.e., $(q_n, p_n) = (q, p)(t = n-0)$. The dynamics reads~\cite{Chirikov:2008}
\begin{equation}
    \begin{cases}
    q_{n+1} = q_n + p_{n+1} \mod 2\pi\,,\\
    p_{n+1} = p_n + K \sin q_n \mod 2\pi\,.
    \end{cases}
\end{equation}
Parameter $K$ controls the dynamics of kicked rotor. For $K\geq K_C \approx 0.972$ \cite{Chirikov:2008}, 
the phase space is roughly separated into chaotic sea and integrable islands. See Fig.~\ref{fig:clkr} for the case of $K = 1.1$. 

\begin{figure}
    \centering
    \includegraphics[width=\linewidth]{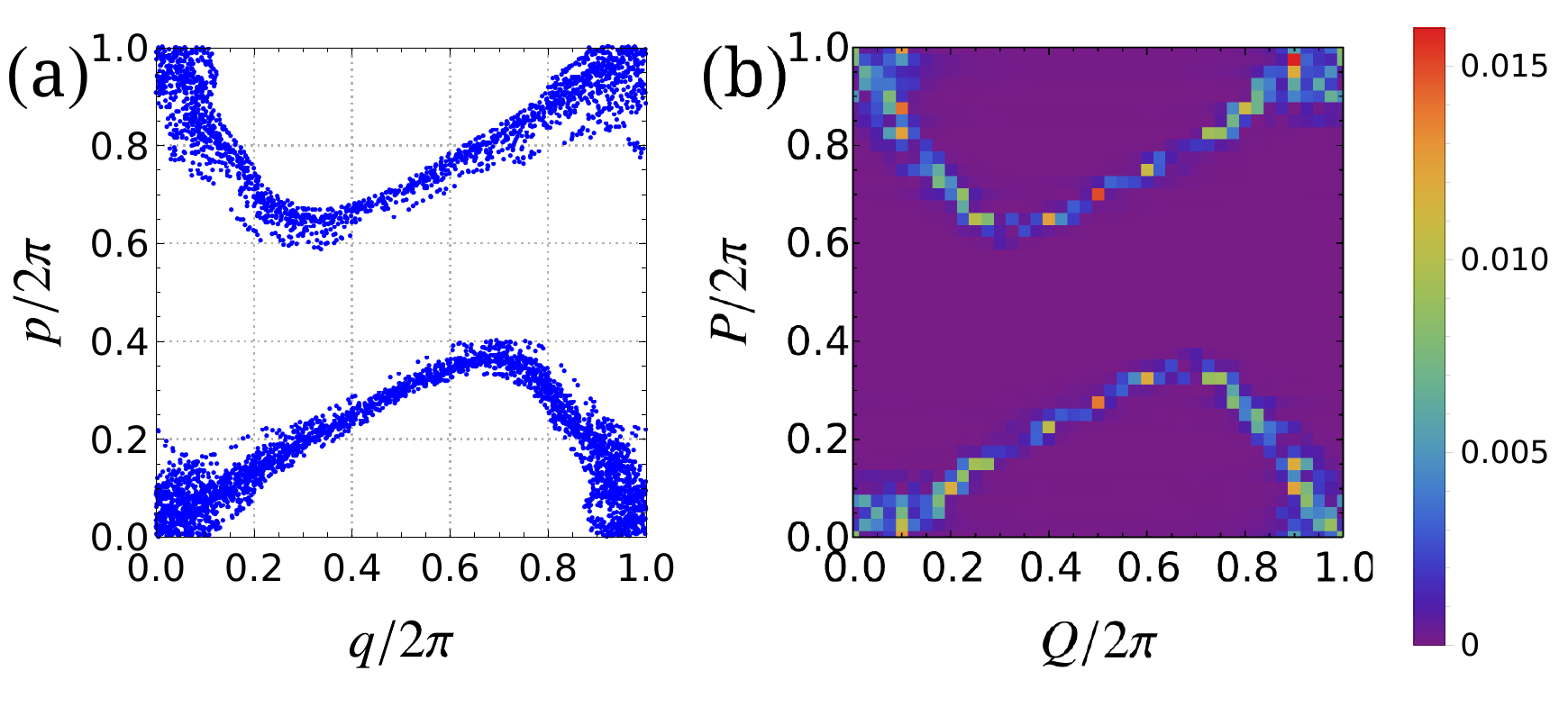}
    \includegraphics[width=\linewidth]{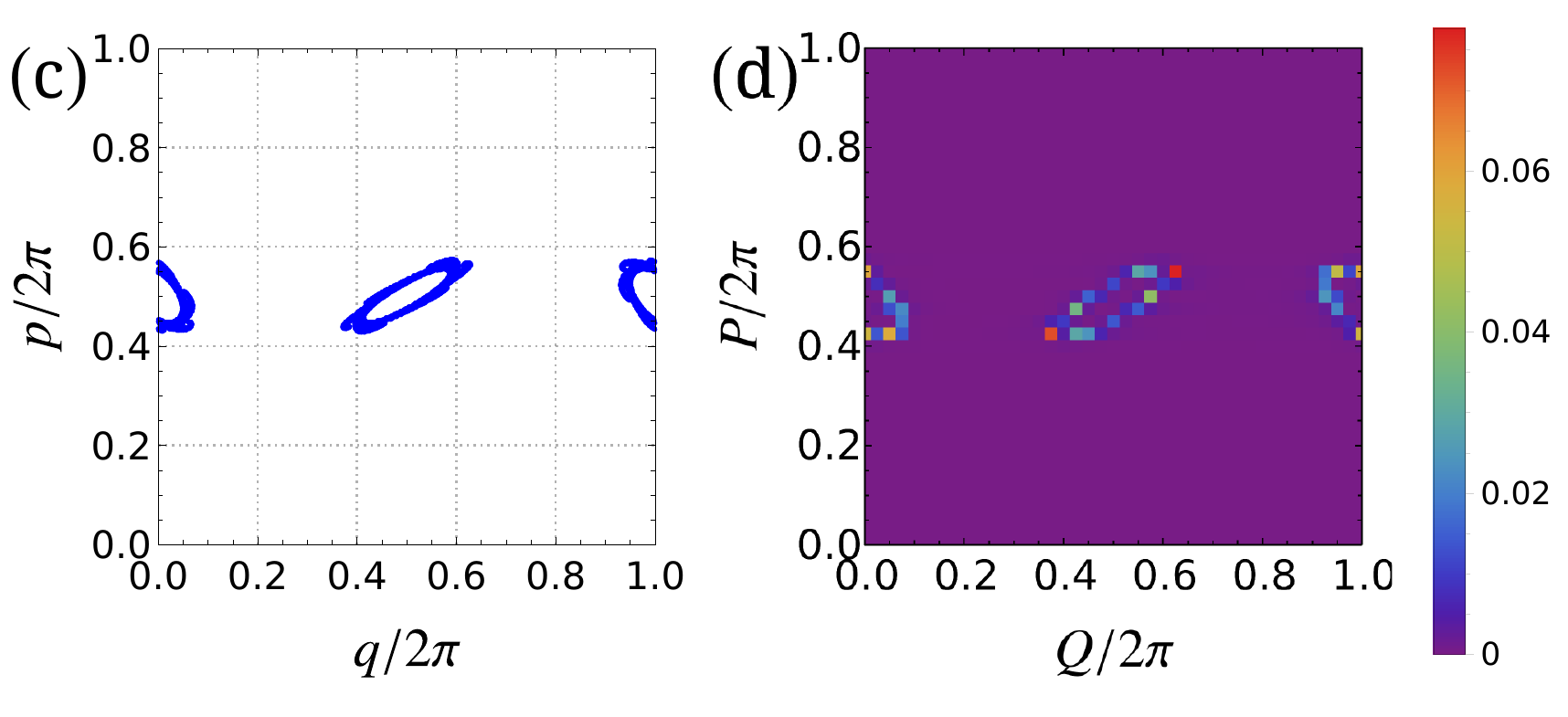}
    \caption{The typical chaotic(a) and integrable(c) trajectories of classical kicked rotor and their corresponding quantum Floquet states (b), (d). The parameter $K=1.1$ and the effective Planck constant is $\hbar_{\rm eff}\approx 0.0039$, i.e., there are $40\times 40$ Planck cells in total.}
    \label{fig:quclcorr_kr}
\end{figure}

As there is a periodic kicking, the quantum dynamics of the kicked rotor is given by the following unitary Floquet evolution
\begin{equation}
    \hat U = e^{- i \hat p^2 / 2\hbar} e^{-i K \cos \hat q/\hbar}.
\end{equation}
The analogs to energy eigenstates and eigenvalues here are the eigenstates of $\hat U$ (Floquet states) and their pseudo-energies \cite{PhysRev.138.B979,PhysRevA.7.2203,Okuniewicz}. For the continuity of notation, we still denote them as $\ket{E}$ and $E$, i.e., $\hat U\ket{E} = e^{-i E} \ket{E}$. We point out that there still holds the correspondence between Floquet states and the the classical invariant distributions as we argued in Sec. \ref{sec:theory}. This is illustrated in Fig.~\ref{fig:quclcorr_kr} with the same setup as 
in Ref.~\cite{otoc, qkam, jljiang2017}. The phase space is discretized into $m\times m$ Planck cells 
and the effective Planck constant is $\hbar_{\textrm{eff}} = 2\pi \hbar / m^2$.

\begin{figure}
    \centering
    \includegraphics[width=\linewidth]{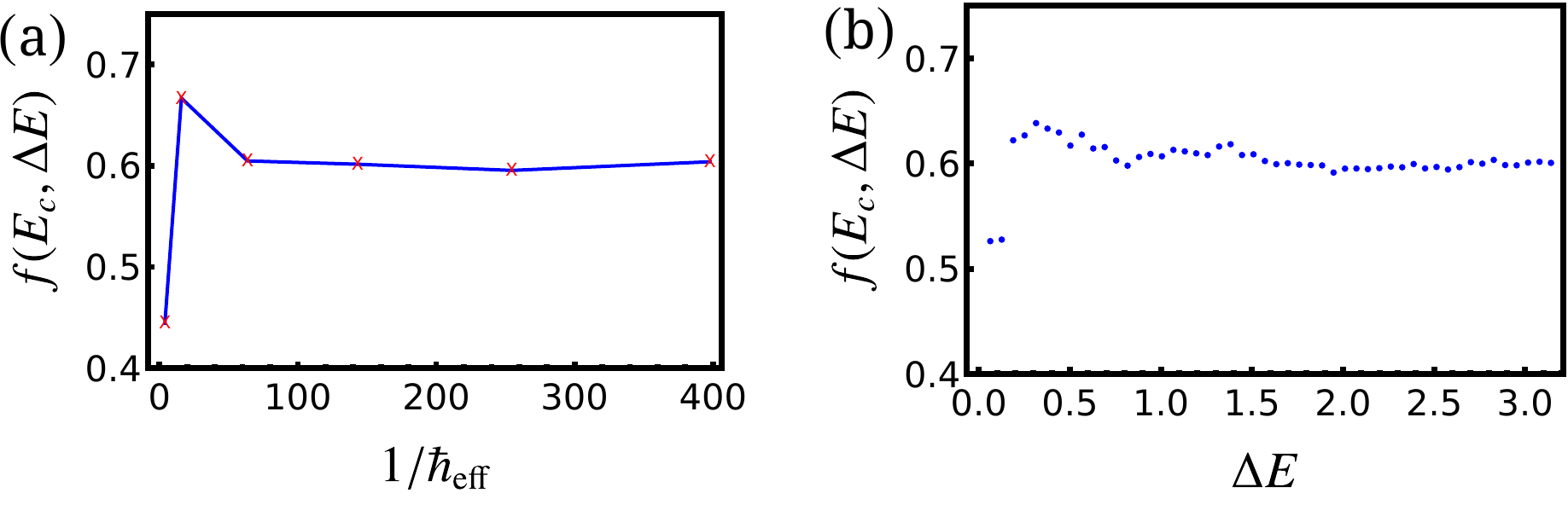}
    \includegraphics[width=\linewidth]{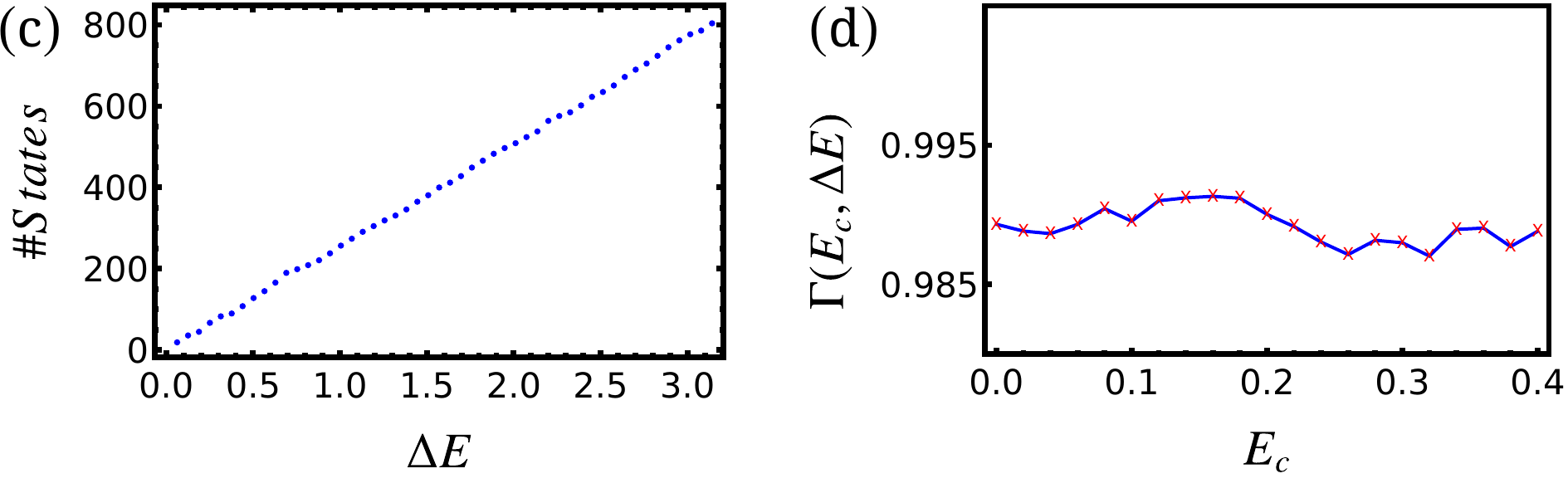}
    \caption{
    The self-similarity with quantum kicked rotor. (a) The ratio $f(E_c,\Delta E)$ as a function of the inverse of the effective Planck constant 
    with $\Delta E = 2$ and $E_c=0$. (b) The ratio as the function of  the width of energy shell $\Delta E$ with $\hbar_{\textrm{eff}}\approx0.0039$ and $E_c=0$. 
    (c) The number of Floquet states as a function of  the width of energy shell $\Delta E$. 
     (d) The dependence of the generalized Wigner-von Neumann entropy $\Gamma(E_c,\Delta E)$ on the center pseudo-energy $E_c$. 
     The pseudo-energy shell width $\Delta E = 0.4$ here.}
    \label{fig:selfsim_kr}
\end{figure}

For this quantum kicked rotor, we discuss the self-similarity with two control parameters of $\hbar_{\textrm{eff}}$ 
and the pseudo-energy shell width $\Delta E$. The self-similarity with $\hbar_{\textrm{eff}}$ is shown in 
Fig.~\ref{fig:selfsim_kr}(a) with the classifier
\begin{equation}
    C_{\textrm{width}}(\ket{E}) = \begin{cases}
    1 & W(\ket{E}) > 0.8 \pi \\
    0 & \textrm{otherwise}
    \end{cases}.
    \label{eq:clasf_width}
\end{equation}
Here $W(\ket{E}) $ is the width of Floquet state in phase space defined with Planck cell basis $\ket{Q,P}$ \cite{otoc}, i.e.,
\begin{equation}
    W(\ket{E})^2 = \sum_{Q, P} \Big[(Q - \pi)^2 + (P-\pi)^2\Big] |\braket{Q,P|E}|^2.
\end{equation}
Geometrically, $W(\ket{E})$ quantifies the ``radius'' of Floquet states to the center point $(q, p) = (\pi, \pi)$. 
With the same classifier, the self-similarity with pseudo-energy shell width $\Delta E$ as the control 
parameter is shown in Fig.~\ref{fig:selfsim_kr}(b), where the ratio $f(E_c,\Delta E)$ quickly saturates 
to a similar value as in Fig.~\ref{fig:selfsim_kr}(a). This demonstrates that decreasing $\hbar_{\textrm{eff}}$ fixing $\Delta E$
is equivalent to increasing $\Delta E$ with fixed $\hbar_{\textrm{eff}}$ as discussed in Appendix \ref{app_deltaE}. 

Although the agreement between the two saturation values in Fig.~\ref{fig:selfsim_kr}(a) and Fig.~\ref{fig:selfsim_kr}(b) 
vindicates our argument in Appendix~\ref{app_deltaE}, we note that our analysis in the Appendix only works 
at the vicinity of $\hbar \sim 0$ and for small $\Delta E$, and the agreement  at such large $\Delta E$ is still surprising. 
This is related to that the pseudo-energies are confined into $[0,2\pi)$ due to the periodic kicking. 
This means that  the different eigen-energies without kicking are folded into this finite interval. 
Thus the kicked rotor has the same dynamical feature at pseudo-energy shells centered at different pseudo-energies, 
similar to the billiard systems where the dynamics is also not sensitive to the energy. 
This is supported by the results in Fig.~\ref{fig:selfsim_kr})(c), where 
the number of eigenstates in the energy shell grows linearly with the shell width. 
This can be further demonstrated by that the entropy of the renormalized distributions of Floquet states in given pseudo-energy 
shells is roughly independent of the center of the shell. Formally, we compute
\begin{equation}
    p_{Q,P}(E_c, \Delta E) = \frac {\sum_{E' \in \mathcal{S}(E_c, \Delta E)} |\braket{Q,P|E'}|^2} {|\mathcal{S}(E_c, \Delta E)|},
\end{equation}
which is actually the probability distribution $p_{Q,P} = \textrm{Tr}[\ket{Q,P}\bra{Q,P} \hat \rho(E_c,\Delta E)]$ of maximally mixed state in the energy shell $\hat \rho(E_c, \Delta E) \propto \sum_{E'\in \mathcal{S}(E_c,\Delta E)} \ket{E'}\bra{E'}$. We then compute 
its generalized Wigner-von Neumann entropy (GWvNE)\cite{PhysRevE.99.052117}  on Planck cell basis (with normalization)
\begin{equation}
    \Gamma(E_c,\Delta E) = -\frac 1 {\log m^2 }\sum_{Q,P} p_{Q,P}(E_c,\Delta E) \log p_{Q,P}(E_c,\Delta E).
    \label{eq:cumu_entropy}
\end{equation}
The function $\log$ is the natural logarithm. By normalization, we make sure $\Gamma \in [0, 1]$. In Fig.~\ref{fig:selfsim_kr}(d), that 
the entropy is almost constant  as the center energy $E_c$ changes demonstrates the independence of dynamics to $E_c$ in the kicked rotor. 
%Thus, by scaling up the thickness of pseudo-energy shell, more Floquet states are counted and tends to reconstruct 
%the whole phase space. This procedure is similar to our argument about isoenergetic surfaces 
%and is same to our discussion about circular billiard. 

\begin{figure*}[t]
    \centering
    \includegraphics[width=0.9\textwidth]{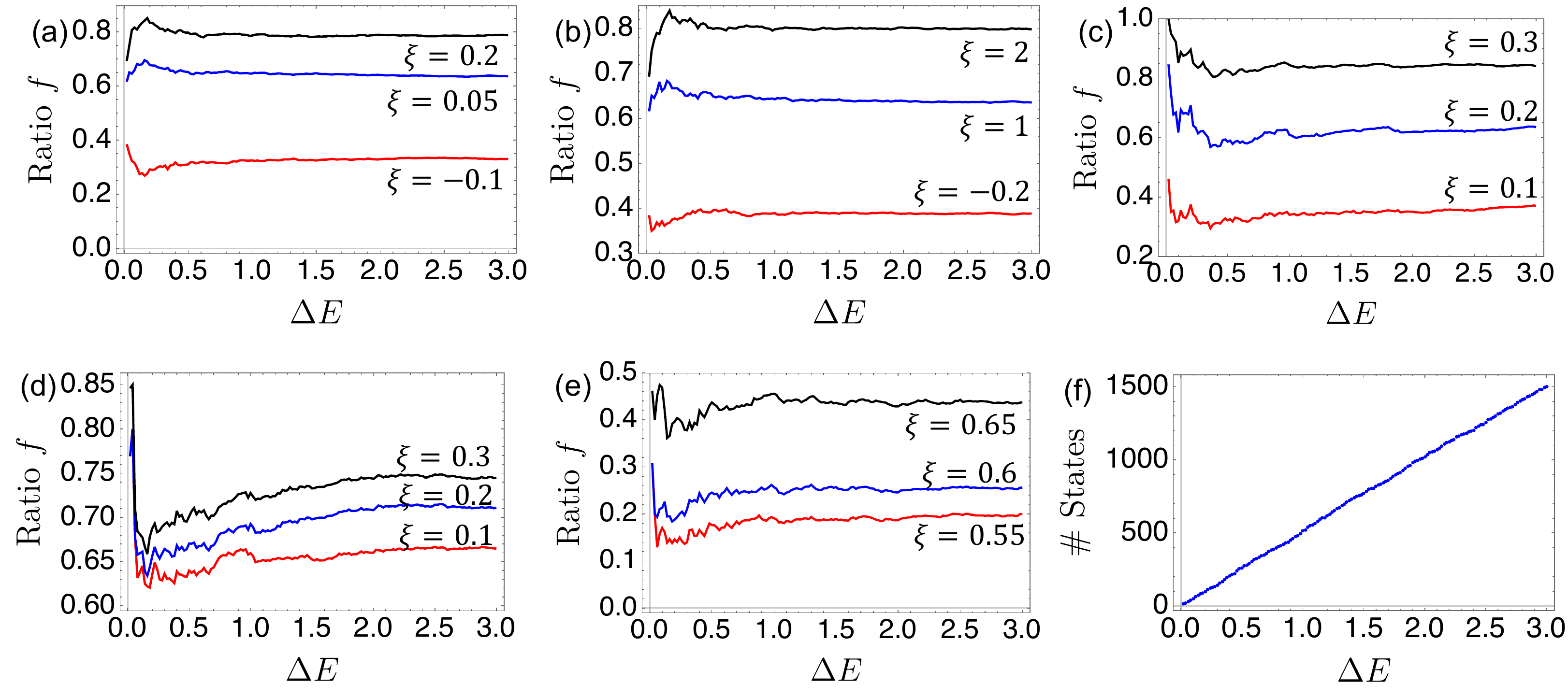}
    \caption{The ratio $f$ for various classifiers: (a)magnetization for a single spin (6th-spin), (b)total magnetization, (c)two-spin-correlation for 3rd spin and 4th spin, (d)global-X-correlation, (e)2nd R\'enyi entropy of 1st spin. (f) shows the total number of energy eigenstates in energy shell $[E_{c}-\Delta E/2,E_{c}+\Delta E/2]$. We take the length of spin chain to be $N=12$, parameter $\rho=0.4$ and central energy $E_{c}=4$.}
    \label{fig:XXZ}
\end{figure*}

\subsection{Heisenberg XXZ Model}
Though our explanation of self-similarity relies on the quantum-classical correspondence, we find that it exits  for quantum systems
that do not have well defined classical limits.  Here is an example, 
one-dimensional Heisenberg XXZ model\cite{XXZ}, whose Hamiltonian with open boundary condition is
\begin{equation}
    H=\sum_{i=1}^{N-1} (\sigma_{i}^{x}\sigma_{i+1}^{x}+\sigma_{i}^{y}\sigma_{i+1}^{y}+\rho \sigma_{i}^{z}\sigma_{i+1}^{z}),
\end{equation}
where $\sigma_{i}^{x,y,z}$ are Pauli operators on the $i$-th site.  We use five different functions $F(\ket{\psi_n})$
to characterize the energy eigenstates $\ket{\psi_n}$. The first four are 
expectations of different operators $O$,  $F(O, \ket{\psi_n}) = \braket{\psi_n|O|\psi_n}$. These 
four different operators are: (i) single spin, here we choose $\sigma_{[N/2]}^{z}$; 
(ii) total magnetization $M_{z}=\sum_{i=1}^{N} \sigma_{i}^{z}$; 
(iii) two-point correlation $C_{34} = \sigma_{3}^{x}\sigma_{4}^{x}$;  
(iv) global-X-correlation product $N_x = \prod_{i=1}^{N} \sigma_{i}^{x}$. 
The fifth function the second-R\'enyi entropy\cite{renyi} of the first spin, i.e.
\begin{equation}
    F(|\psi_{n}\rangle) = -\log[\operatorname{Tr}(\rho_{1}^{2})],
\end{equation}
where $\rho_{1}=\operatorname{Tr}_{2,\dots,N}|\psi_{n}\rangle\langle\psi_{n}|$ is the reduced density matrix for the first spin.
For each of these five different functions,  the classifier is defined as 
\begin{equation}
C(|\psi_n\rangle; \xi) = \begin{cases}
1 & F(\ket{\psi_n}) \geq \xi\,, \\
0 & \textrm{otherwise}\,,
\end{cases}
\end{equation}
where $\xi$ is a chosen threshold value. 

All calculations are done by exact diagonalization and the results are shown in Fig.~\ref{fig:XXZ}.
For all the local operators \cite{k-local}, the total magnetization, single spin and two-point-correlation operator, the ratio functions $f(\Delta E)$
quickly saturate as $\Delta E$, indicating the self-similarity. The result for the the second-R\'enyi entropy is similar.
However,  for the global-X-correlation, the ratio does not saturate  at a constant value even if there are hundreds of energy eigenstates 
in the energy shell. It seems that  the  self-similarity fails for non-local operators.

Our results can be interpreted in the context of observability of operators. When we observe a certain property of Heisenberg-XXZ spin 
chain in a macroscopic way,  local properties, such as magnetization, correlation or reduced density matrix for one spin, can be measured 
with macroscopic instruments. However, non-local operators such as global-X-correlation, do not have classical macroscopic meaning 
and their properties can not be measured with a classical instrument.  

%for example, observing the total magnetization, two-point-correlation, entanglement, single spin magnetization and global-X-correlation, rather than measuring those observables for a single energy eigenstate, the mean of energy eigenstates in an energy shell $[E_{c}-\Delta E/2,E_{c}+\Delta E/2]$ will determine the properties of spin chain. 

\section{Conclusion and Discussion}
We have discovered self-similarity among energy eigenstates of a quantum system. If all the eigenstates are ordered according
to their corresponding eigenvalues and are divided 
into different groups according to their properties or features, then the members of the same group tend to ``repel" each other
and scatter rather evenly among all the eigenstates. In other words, inside any energy shell  ${\mathcal S}(E_c,\Delta E)$, 
the composition of eigenstates from different groups is very similar. Just like a shoreline, any piece looks similar to the other piece
as long as the piece is not too short. We have offered an explanation why this self-similarity should exist in 
quantum systems that have well-defined classical limits. Our numerical results show that this self-similarity also exists in 
quantum systems that have no well-defined classical limits. 

The three quantum systems that we studied in Sec. III have a common feature: their Hilbert spaces have finite dimensions. 
When we enlarge their Hilbert spaces, for example, by increasing $j$ (or $N$) or adding more spins,  each of these quantum systems
will have more eigenstates. The self-similarity suggests that despite that  the Hilbert space is getting bigger, the composition 
of eigenstates will remain roughly the same. It is analogue to shuffled cards: when you insert one deck of cards randomly into, 
one hundred decks of thoroughly shuffled  cards, nothing changes significantly besides there are more cards. 
One may use this to define the classical limit of a quantum system, in particular, the quantum system that has no well defined
classical counterpart: when the composition of different types of eigenstates no longer changes with the dimension of the Hilbert space, 
the quantum system reaches its classical limit.

Modern statistical mechanics has a very basic hypothesis, the postulate of equal a priori probability\cite{huang}. 
It is regarded as a working hypothesis, that is, its justification comes not from that it is derived from 
the fundamental microscopic theory but from the fact that the conclusions derived from it agree with experiments\cite{huang}. 
The self-similarity among  energy eigenstates offers a reasonable  justification for this hypothesis from a microscopic perspective. 
According to this postulate, for a quantum system, the micro-canonical ensemble average of any physical observable $\hat{O}_{\mathrm{phy}}$ should be $\braket{\hat{O}_{\mathrm{phy}}}_{\Delta E}=\operatorname{Tr}[\hat{O}_{\mathrm{phy}}\hat{\rho}_{\Delta E}]$ with 
\begin{equation}\label{eq:microcanonical}
    \hat{\rho}_{\Delta E}=\frac{1}{\Omega(E_{c},\Delta E)}\left[\sum_{E\in\mathcal{S}}|E\rangle\langle E|\right]\,,
\end{equation}
where $\Omega(E_{c},\Delta E)$ is the total number of energy eigenstates in the energy shell $\mathcal{S}(E_{c},\Delta E)$.
Let us project every eigenstate $\ket{E}$ into the phase space, then the probability of the system at the Planck cell $|Q,P\rangle$ is
\begin{equation}
\text{Prob}(Q, P)=\frac{1}{\Omega(E_{c},\Delta E)}\left[\sum_{E\in\mathcal{S}}|\langle Q,P|E\rangle|^2\right]\,,
\end{equation}
where $Q,P$ are multi-dimensional variables. The self-similarity implies that this probability is zero outside 
of the shell $\mathcal{S}(E_{c},\Delta E)$ and the same for every Planck cell in the shell. 
This holds for any quantum system, integrable or chaotic. For a fully chaotic system, every eigenstate $|E\rangle$ 
inside the shell looks about the same; consequently, the density matrix (\ref{eq:microcanonical}) is still valid even 
as the width $\Delta E$ is so small that there is only one eigenstate in the shell. This is just the well known eigenstate thermalization 
hypothesis\cite{Landau1980,Srednicki1994}. For systems that are not fully chaotic, the width $\Delta E$ has to be large enough so that different types of eigenstates
are properly represented statistically. Overall, this explains why there is no need to consider the integrability of the system 
when the micro-canonical density matrix (\ref{eq:microcanonical}) is used.

\section{Acknowledgement}
 This work is supported by the National Key R\&D Program of China (Grant No.~2018YFA0305602), National Natural Science Foundation of China (Grant No. 11921005), and Shanghai Municipal Science and Technology Major Project (Grant No.2019SHZDZX01).
 
%\bibliography{apssamp}

\providecommand{\noopsort}[1]{}\providecommand{\singleletter}[1]{#1}%

\begin{appendix}

\section{Relation Between Decreasing Planck Constant and Enlarging Energy Window}\label{app_deltaE}
When discussing the circular billiard, we have used the fact that the decreasing of $\hbar$ is equivalent 
to simply increasing the shell width $\Delta E$ with fixed $E_c$. 
This is clearly right for the case of billiards; for a general case, it needs some more clarification.

Let us discuss the general case and the formal definition of the eigentstates in the energy shell here. 
Consider a general quantum system near the limit $\hbar\rightarrow 0$, the energy levels in the interval of $\mathcal{S}(E_c, \Delta E)$ 
could be expressed by a quantum number relative to $E_c$, i.e., by letting $E_c = E_{n_c}(\hbar)$, 
then $E_n(\hbar) = E_{n_c + (n-n_c)}(\hbar) \approx E_c + (n-n_c)^\beta\hbar^\alpha\equiv E_c + \hbar^\alpha G_{n-n_c}(E_c)$. 
Here $\alpha > 0$ and $G_n$ is some function independent of $\hbar$. We note that the nearest energy spacing in a fixed energy 
shell  $\mathcal{S}(E_c, \Delta E)$ should vanish as $\hbar\rightarrow 0$, thus we always have the factor of $\hbar^\alpha$. 
Here are two examples.  For a one-dimensional billiard, we have $E_n \propto \hbar^2 n^2\Rightarrow E_n 
\approx E_c + 2E_c\hbar^2 (n-n_c)$ with $\alpha=2$. And for systems similar to hydrogen atoms, 
we have  $E_n \propto 1 / \hbar^2 n^2$, and thus $E_n \approx E_c - 2 (n-n_c) \hbar / E_c^{3/2}$, which means $\alpha = 1$.

With the above consideration, the collection of eigenstates within a given energy shell can then be  defined as 
\begin{equation}
\begin{aligned}
 \Upsilon(E_c, \Delta E, \hbar) &= \{\psi_{n, \hbar}: E_n(\hbar) \in \mathcal{S}(E_c, \Delta E)\} \\
 &=\{\psi_{n,\hbar}: \hbar^\alpha G_n \in \mathcal{S}(0, \Delta E)\}
\end{aligned}
\end{equation}
With $\hbar$ decreasing by a factor of $w>1$, we 
have $\Upsilon(E_c, \Delta E, \hbar / w) = \{\psi_{n, \hbar / w}, \hbar^\alpha G_n / w^\alpha \in \mathcal{S}(0, \Delta E)\} = \{\psi_{n, \hbar / w}, \hbar^\alpha G_n \in \mathcal{S}(0, w^\alpha \Delta E)\}$. If the physics of eigenstates of fixed level is not sensitive to $\hbar$ (this is true for quantum systems 
who have clear classical counterparts, since eigenstates become classical invariant distributions in phase space as $\hbar \rightarrow 0$), 
we have the asymptotic behavior for the collection of eigenstates,
\begin{equation}
    \Upsilon(E_c, \Delta E, \hbar / w) = \Upsilon(E_c, w^\alpha \Delta E, \hbar).
    \label{eq:app_A_asym}
\end{equation}
This means that changing $\hbar$ and changing $\Delta E$ at the opposite directions are the same on the population of eigenstates, 
so does it on any statistical properties of eigenstates, e.g., the self-similarity.

This equivalence of increasing $\Delta E$ and decreasing $\hbar$) relies on two assumptions. (1) 
The expansion of $E_{n_c+(n-n_c)} \approx E_c +\hbar^\alpha G_n$ near $E_c$ has a high precision; 
(2) the eigenstates are not sensitive to $\hbar$. 
For billiards, both of these two assumptions are satisfied exactly. The energy levels has a natural cut-off at $\alpha=2$ while the eigenfunction is actually the solution of $(\nabla^2 + k_{nm}^2)\psi_{nm} = 0$ which is independent of $\hbar$. For other systems discussed in the main text, these two assumptions 
are not exactly satisfied. For the double-top and kicked rotor, the equivalence still works well as their classical counterparts are  well-defined. 
For the XXZ model, we should expect that there is difference between enlarging $\Delta E$ and reducing $\hbar$. 

We also note that the center of energy shell is the same on both sides of Eq. (\ref{eq:app_A_asym}). This holds when 
$E_c = E_{n_c}$ has a solution. In general, we can only choose $n_c$ as the closest quantum energy level near $E_c$ and 
this would make the right hand side of Eq. (\ref{eq:app_A_asym}) has a different $E_c$. However, we expect when $\hbar$ goes to zero, this shifting would vanish since the energy levels become sufficient dense and $E_c = E_{n_c}$ would has a solution of arbitrarily high precision.

\section{Classical Correspondence of Quantum Coupled Top}\label{app_pathintegral}
\begin{figure*}[t]
    \centering
    \includegraphics[width=0.9\textwidth]{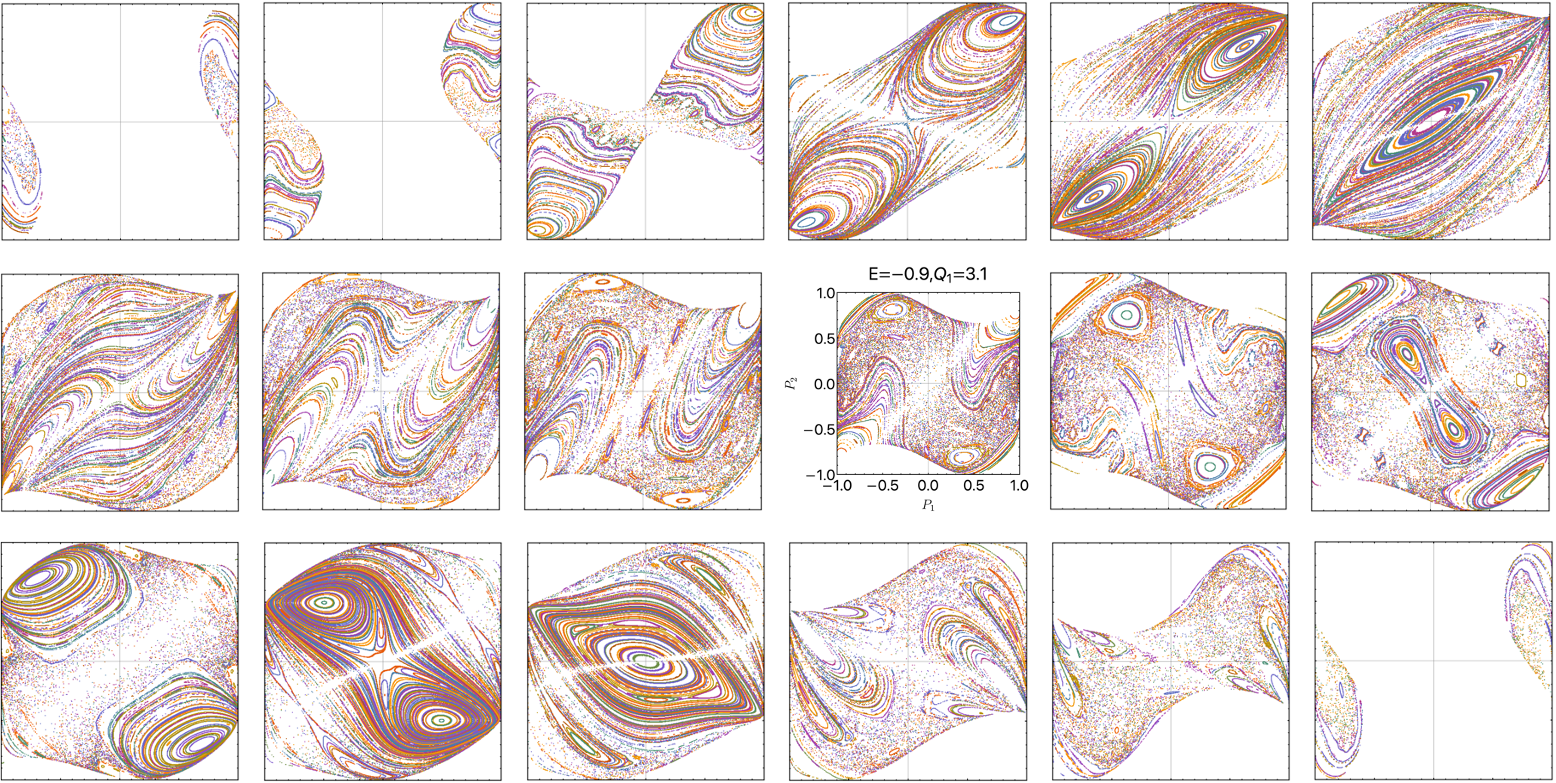}
    \caption{The coupled-top's classical \poincare sections of various $Q_{1}$ with $E=-0.9$ fixed. The partition we take is $\{$0.1, 0.7, 1.0, 1.3, 1.6, 1.9, 2.2, 2.5, 2.8, 3.1, 3.4, 3.7, 4.0, 4.3, 4.6, 4.9, 5.2, 5.8$\}$ (from left to right, up to down).}
    \label{fig:app_toppsec}
\end{figure*}
We start with Hamiltonian (\ref{eq:doubletopH}), and use path integral method to rigorously derive the system's classical counterpart. 
We first review some basics of  coherent states of spin-$J$, which are defined as 
\begin{equation}
\begin{aligned}
    \left|\mu\right\rangle=&\frac{1}{\left(1+|\mu|^{2}\right)^{J}}\times \\
    &\sum_{n=0}^{2 J} \sqrt{\frac{(2 J) !}{(2 J-n) ! n !}}\, \mu^{n}\left|L_{z}=J-n\right\rangle\,,
\end{aligned}
\end{equation}
where $\mu=e^{\mathrm{i} \phi} \tan(\theta/2)$. We have following properties for spin coherent states,
\begin{align}
    &\left\langle\mu^{\prime} | \mu\right\rangle =\frac{\left(1+\mu^{\prime *} \mu\right)^{2 J}}{(1+\left|\mu^{\prime}\right|^{2})^{J}\left(1+|\mu|^{2}\right)^{J}}, \\
    &\langle\mu^{\prime}|\hat{L}_{+}| \mu\rangle =\frac{2 J \mu}{1+\mu^{\prime *} \mu}\left\langle\mu^{\prime} | \mu\right\rangle, \\
    &\langle\mu^{\prime}|\hat{L}_{z}| \mu\rangle =J \frac{1-\mu^{\prime *} \mu}{1+\mu^{*} \mu}\left\langle\mu^{\prime} | \mu\right\rangle, \\
    &1 =\int \frac{2J+1}{4\pi}\mathrm{d}\Omega |\mu\rangle\langle\mu|. \label{eq:closure}
\end{align}
We want to evaluate  the propagator $ \langle \boldsymbol{\mu}(t')|e^{-\mathrm{i} \hat{H} T}| \boldsymbol{\mu}(t)\rangle$ with $T=t'-t$. Here $\boldsymbol{\mu}\equiv(\mu_{1},\mu_{2})$, $|\boldsymbol{\mu}\rangle\equiv|\mu_{1}\rangle\otimes|\mu_{2}\rangle$, which corresponds a double-spin coherent state.
We  divide the time interval $T$ into $N+1$ equal pieces of duration $\delta t=T/(N+1)$, and denote 
$\boldsymbol{\mu}_i=\boldsymbol{\mu}(t)+i\delta t$. After inserting $N$ coherent-states' closure-relations (\ref{eq:closure}), we get
\begin{equation}
\begin{aligned}
    &\langle \boldsymbol{\mu}(t')|e^{-\mathrm{i} \hat{H} T}| \boldsymbol{\mu}(t)\rangle\\
    =& \int \prod_{i=1}^{N} \left(\frac{2 J+1}{4 \pi}\right)^{2} \mathrm{d}\Omega_{1,i}\mathrm{d}\Omega_{2,i} \langle\boldsymbol{\mu}_{i+1}|e^{-\mathrm{i} \hat{H} \delta t}| \boldsymbol{\mu}_{i}\rangle \\
    =& \int \mathcal{D}[\mu] \prod_{i=1}^{N}\left(\langle\boldsymbol{\mu}_{i+1} | \boldsymbol{\mu}_{i}\rangle-\mathrm{i}\langle\boldsymbol{\mu}_{i+1}|\hat{H}| \boldsymbol{\mu}_{i}\rangle \delta t+\mathcal{O}\left(\delta t^{2}\right)\right) \\
    =& \int \mathcal{D}[\mu] \prod_{i=1}^{N}\langle\boldsymbol{\mu}_{i+1} | \boldsymbol{\mu}_{i}\rangle\bigg[1-\mathrm{i} \delta t\bigg(-\frac{2 J \operatorname{Re} \mu_{1,i}}{1+\mu_{1, i+1}^{*} \mu_{1, i}}\\
    &-(1 \leftrightarrow 2)-\mu J \frac{1-\mu_{1, i+1}^{*} \mu_{1, i}}{1+\mu_{1, i+1}^{*} \mu_{1, i}} \frac{1-\mu_{2, i+1}^{*} \mu_{2, i}}{1+\mu_{2, i+1}^{*} \mu_{2, i}}\bigg)\bigg]\\
    =& \int \mathcal{D}[\mu] \exp(\mathrm{i}S[\boldsymbol{\mu}(t),\boldsymbol{\mu}(t')])
\end{aligned}
\end{equation}
Here $\mathcal{D}[\mu]\equiv\prod_{i=1}^{N} \left(\frac{2 J+1}{4 \pi}\right)^{2} \mathrm{d}\Omega_{1,i}\mathrm{d}\Omega_{2,i}$ is the functional integration measure, and $\mathrm{d}\Omega_{1,i}$ ($\mathrm{d}\Omega_{2,i}$) is the solid angle measure of the first (second) top at lattice $\boldsymbol{\mu}_{i}$. The discrete form of action immediately reads
\begin{equation}
\begin{aligned}
    \frac{S[\boldsymbol{\mu},\boldsymbol{\mu}']}{2 J}=& \sum_{i=1}^{N}-\mathrm{i} \log \prod_{\alpha=1,2} \frac{1+\mu_{\alpha, i+1}^{*} \mu_{\alpha, i}}{\sqrt{(1+|\mu_{\alpha}^{\prime}|^{2})(1+|\mu_{\alpha}|^{2})}}\\
    &+\delta t\bigg(\frac{\operatorname{Re} \mu_{1, i}}{1+\mu_{1, i+1}^{*} \mu_{1, i}}+\frac{\operatorname{Re} \mu_{2, i}}{1+\mu_{2, i+1}^{*} \mu_{2, i}}\\
    &+\frac{\mu}{2} \frac{1-\mu_{1, i+1}^{*} \mu_{1, i}}{1+\mu_{1, i+1}^{*} \mu_{1, i}} \frac{1-\mu_{2, i+1}^{*} \mu_{2, i}}{1+\mu_{2, i+1}^{*} \mu_{2, i}}\bigg).
\end{aligned}
\end{equation}
In the continuous limit, substitute $\mu_{i+1}=\mu_{i}+\mathrm{d}\mu$, we get
\begin{equation}
    \frac{1+\mu^{*} \mu}{\sqrt{(1+|\mu^{\prime}|^{2})(1+|\mu|^{2})}}=1+\frac{1}{2} \frac{\mu \mathrm{d} \mu^{*}-\mu^{*} \mathrm{d} \mu}{1+|\mu|^{2}}.
\end{equation}
Then, the action can be written as the functional
\begin{equation}
\begin{aligned}
    \frac{\mathcal{S}[\boldsymbol{\mu},\boldsymbol{\mu}']}{J}&=\int \mathrm{d} t\bigg[-\mathrm{i} \frac{\mu_{1} \dot{\mu}_{1}^{*}-\mu_{1}^{*} \dot{\mu}_{1}}{1+\left|\mu_{1}\right|^{2}}-\mathrm{i} \frac{\mu_{2} \dot{\mu}_{2}^{*}-\mu_{2}^{*} \dot{\mu}_{2}}{1+\left|\mu_{2}\right|^{2}}\\
    &+\cos \phi_{1} \sin \theta_{1}+\cos \phi_{2} \sin \theta_{2}+\mu \cos \theta_{1} \cos \theta_{2}\bigg].
\end{aligned}
\end{equation}
The (normalized) Lagrangian reads
\begin{equation}
\begin{aligned}
    \mathcal{L}=&-\left(1-\cos \theta_{1}\right) \dot{\phi}_{1}-\left(1-\cos \theta_{2}\right) \dot{\phi}_{2}\\
    &+\cos \phi_{1} \sin \theta_{1}+\cos \phi_{2} \sin \theta_{2}+\mu \cos \theta_{1} \cos \theta_{2}.
\end{aligned}
\end{equation}
In the classical limit, i.e., $J\rightarrow +\infty$ (or equivalently, $\hbar\rightarrow 0$), the propagator is permutation matrix with the jumping governed by the classical canonical equation of motion \ref{eq:doubletop_eom}.

\section{Angular Momentum Planck Cells}\label{app_planckcell}
In this section of appendix, we give the proof of properties regarded to Planck cells \ref{eq:ketQP}. Recall the definition of Planck cells\cite{otoc}:
\begin{equation}
    |Q, P\rangle=\frac{1}{\sqrt{\Delta_{p}}} \int_{P}^{P+\Delta_{p}} d p|p\rangle e^{-i Q p / \hbar},
\end{equation}
where $\Delta_{q}$ and $\Delta_{p}$ are size of cells along $q,p$ axis with $\Delta_{q}\Delta_{p}=2\pi\hbar$.
Express it in a discrete formalism, we obtain the Planck cells for a single angular momentum:
\begin{equation}
    \vert Q,P\rangle=\frac{1}{\sqrt{L}}\sum_{n=-m+Pj}^{m+Pj} e^{-iQn} \vert L_{z}=n\rangle,
\end{equation}
which is \ref{eq:ketQP}. It is unambiguous that the $\hat{L}_{z}$ has mean value approximate to $Pj$ while variation is in a same magnitude with $m$, i.e. proportional to $1/\sqrt{\hbar}$. In the classical limit, this form of Planck cell is equivalent to\cite{otoc}
\begin{equation}
    \vert Q,P\rangle^{(\mathrm{position})}=\frac{1}{\sqrt{\Delta_{q}}} \int_{Q}^{Q+\Delta_{q}} d q|q\rangle e^{i P q / \hbar},
\end{equation}
which means the $\hat{Q}$ also has mean value approximate to its classical correspondence with variation proportional to $1/\sqrt{\hbar}$. The above analysis can also be done by brute force calculation using algebraic relations of quantum angular momentum.

\section{Coupled-Top's Phase Space Volume}\label{app_volume}
\begin{figure}
    \centering
    \includegraphics[width=1.0\linewidth]{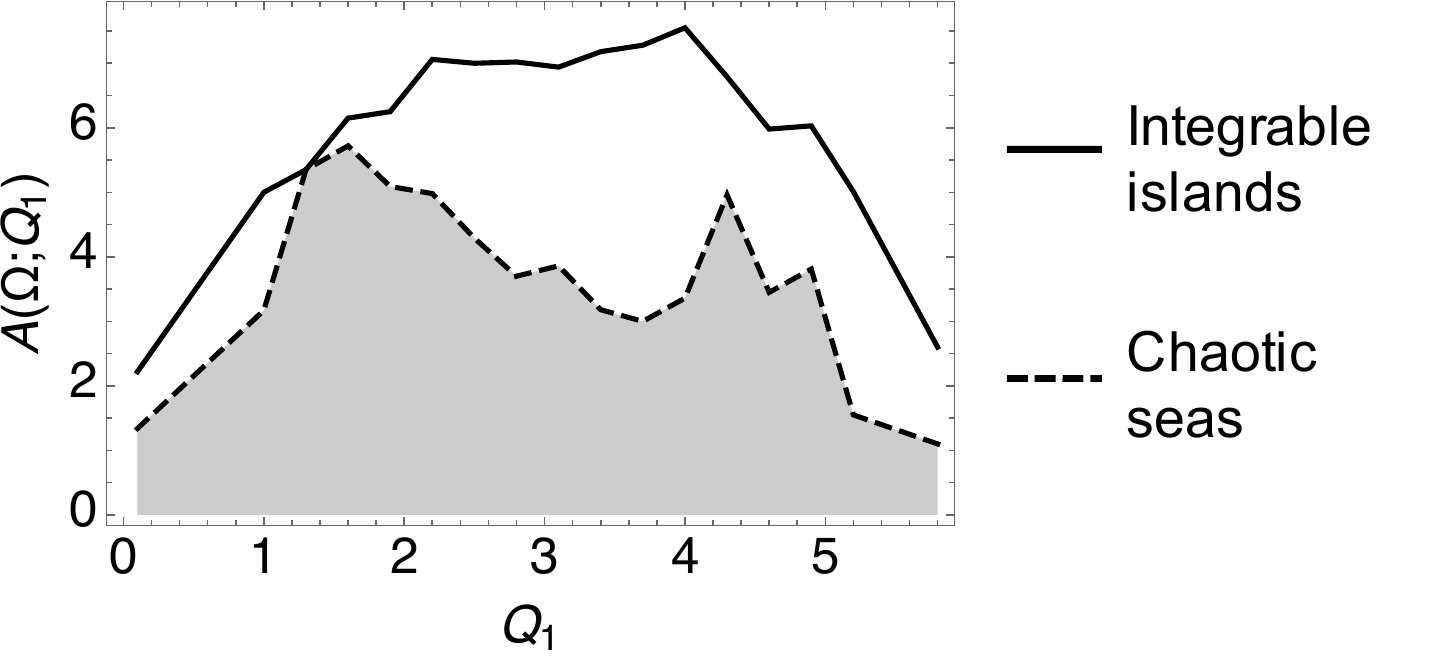}
    \caption{Plot of integral $A(\Omega;Q_{1})$. As for black line, $\Omega$ represents `total isoenergetic surface'; while for dashed line $\Omega$ denotes integrable islands. }
    \label{fig:AOmegaQ1}
\end{figure}

Coupled-top's phase space is a 4-dimensional manifold. Therefor the integrable islands and chaotic seas on a specific isoenergetic surface are 3-dimensional hypersurface. However, only sections of isoenergetic surface are visible to us. In this part of appendix, we show how to calculate the area of integrable islands and chaotic seas via plotting \poincare sections. The overall integration measure for evaluating area of 3-dimensional hypersurface $\Omega$ is
\begin{equation}
    V(\Omega)=\int_{0}^{2\pi} dQ_{1} A(\Omega;Q_{1})\,,
\end{equation}
where 
\begin{equation}
\begin{aligned}
    &A(\Omega;Q_{1})\\
    =&\iint_{\mathbf{P}\in \Omega} dP_{1} dP_{2}
    C_{1}(E;Q_{1},P_{1},P_{2})C_{2}(E;Q_{1},P_{1},P_{2}).
\end{aligned}
\end{equation}
with two projection factors $C_{1}$ and $C_{2}$ defined as
\begin{align}
    &C_{1}(E;Q_{1},P_{1},P_{2})=\sqrt{1+\left(\frac{\partial Q_{2}}{\partial P_{1}}\right)^{2}+\left(\frac{\partial Q_{2}}{\partial P_{2}}\right)^{2}},\\
    &C_{2}(E;Q_{1},P_{1},P_{2})=\frac{1}{\sqrt{1-\left(\frac{\partial H/\partial Q_{1}}{|\nabla H|}\right)^{2}}}.
\end{align}
Here $Q_{2}$ is interpreted as a function of $(E;Q_{1},P_{1},P_{2})$, see Eq. (\ref{eq:Q2}); and $H$ is interpreted as a function of $(Q_{1},Q_{2},P_{1},P_{2})$ with $Q_{2}$ evaluated via Eq.(\ref{eq:Q2}) after partial derivative.
The 3-dimensional area of region $\Omega$ can be approximated as a summation
\begin{equation}
    V(\Omega)\approx \sum_{i=1}^{N-1}\frac{A(\Omega,Q_{1,i})+A(\Omega,Q_{1,i+1})}{2}\Delta Q_{1,i},
\end{equation}
where $\{Q_{1,1},\dots,Q_{1,N}\}$ is a partition of $[0,2\pi]$. The two-dimensional integral $A(\Omega;Q_{1})$ is done in each \poincare section numerically, and the results are plotted in Fig.\ref{fig:AOmegaQ1}.

\end{appendix}

\end{document}